\documentclass[twocolumn,10pt]{./asme2ej}

\usepackage{amsmath}
\usepackage{bm}
\usepackage{dsfont}
\usepackage{graphicx,color,url}
\usepackage{multirow}
\usepackage{chngpage}
\usepackage{array}
\usepackage{float}

\title{Dynamic Response Optimization of Complex Multibody Systems in a Penalty Formulation using Adjoint Sensitivity }

\author{Yitao Zhu \\
    {\tensfb Daniel Dopico } 
	\affiliation{
	\hspace{-20 pt}Advanced Vehicle Dynamics Laboratory \\ \hspace{-20 pt}
	and Computational Science Laboratory\\ \hspace{-20 pt}
	Department of Mechanical Engineering \\ \hspace{-20 pt}
	and Department of Computer Science\\ \hspace{-20 pt}
	Virginia Tech\\
	Blacksburg, VA 24061, USA \\
    Email: yitao7@vt.edu,
    Email: ddopico@vt.edu
    }	
}

\author{Corina Sandu  \thanks{Address all correspondence to this author.}\\
    \affiliation{
    Advanced Vehicle Dynamics Laboratory \\
    Department of Mechanical Engineering \\
	Virginia Tech\\
	Blacksburg, VA 24061, USA \\
	Email: csandu@vt.edu
    }
}

\author{Adrian Sandu
    \affiliation{
	Computational Science Laboratory \\ 
        Department of Computer Science\\
	Virginia Tech\\
	Blacksburg, VA 24061, USA \\
	Email: sandu@cs.vt.edu
    }
}

\begin{document}

\maketitle

\begin{abstract}
{\it Multibody dynamics simulations are currently widely accepted as valuable means for  dynamic performance analysis of mechanical systems. The evolution of theoretical and computational aspects of the multibody dynamics discipline make it conducive these days for other types of applications, in addition to pure simulations. One very important such application is design optimization. A very important first step towards design optimization is sensitivity analysis of multibody system dynamics. Dynamic sensitivities are often calculated by means of finite differences. Depending of the number of parameters involved, this procedure can be computationally expensive. Moreover, in many cases, the results suffer from low accuracy when real perturbations are used. The main contribution to the state-of-the-art brought by this study is the development of the adjoint sensitivity approach of multibody systems in the context of the penalty formulation. The theory developed is demonstrated on one academic case study, a five-bar mechanism, and on one real-life system, a 14-\textit{DOF} vehicle model. The five-bar mechanism is used to illustrate the sensitivity approach derived in this paper. The full vehicle model is used to demonstrate the capability of the new approach developed to perform sensitivity analysis and gradient-based optimization for large and complex multibody systems with respect to multiple design parameters.}
\end{abstract}

\begin{nomenclature}
\entry{\textit{DOF}}{Degree or degrees of a freedom.}
\entry{\textit{DAE}}{Differential algebraic equations.}
\entry{\textit{ODE}}{Ordinary differential equations.}
\entry{\textit{CG}}{Center of gravity.}
\entry{$t $}{Time.}
\entry{$\left(\cdots\right)_0$}{Means evaluation at the initial time $\left(\cdots\right)\left(t_0\right)$.}
\entry{$\left(\cdots\right)_F$}{Means evaluation at the final time $\left(\cdots\right)\left(t_F\right)$.}
\entry {${\bf q} \in \mathds{R}^{n}$}{Vector of coordinates of the system.}
\entry {${\bm \rho} \in \mathds{R}^{p}$}{Vector of parameters.}
\entry{$\left(\right)_{\bf q}$}{$ = \displaystyle \dfrac{\partial \left(\right)}{\partial {\bf q}};
\; \left(\right)_{\bm \rho} = \displaystyle \dfrac{\partial \left(\right)}{\partial {\bm \rho}} $}
\entry{$\dot{\left(\right)}$}{$= \displaystyle \dfrac{{\rm d}\left(\right)}{{\rm d}t};
\; \ddot{\left(\right)} = \displaystyle \dfrac{{\rm d}^2\left(\right)}{{\rm d}t^2}
\; \left(\right)_{t} = \displaystyle \dfrac{\partial \left(\right)}{\partial t}$}
\entry {${\bf M}\left({\bf q},{\bm \rho}\right) \in \mathds{R}^{n \times n}$}{Generalized mass matrix of the system.}
\entry {${\bf Q}\left({\bf q},\dot{\bf q},t,{\bm \rho} \right) \in \mathds{R}^{n}$}{Vector of generalized forces of the system.}
\entry {${\bm \Phi}\left({\bf q},t,{\bm \rho} \right) \in \mathds{R}^{m}$}{Vector of constraints that relate the dependent coordinates.}
\entry{${\bf A}_{\bf x}$}{$ =
\left[\begin{array}{c c c c c}
\displaystyle\frac{\partial{\bf A}}{\partial x_1} & \ldots &
\displaystyle\frac{\partial{\bf A}}{\partial x_i} & \ldots &
\displaystyle\frac{\partial{\bf A}}{\partial x_{s}}
\end{array}\right] \in \mathds{R}^{q{\times}r{\times}s} $. Third order tensor of derivatives of matrix ${\bf A} \in \mathds{R}^{q{\times}r}$ w.r.t. vector ${\bf x} \in \mathds{R}^{s}$.}
\entry{${\bf A}_{\bf x}^{\rm T}$}{$ =\left[\begin{array}{c c c c c}
\displaystyle\frac{\partial{\bf A}^{\rm T}}{\partial x_1} & \ldots &
\displaystyle\frac{\partial{\bf A}^{\rm T}}{\partial x_i} & \ldots &
\displaystyle\frac{\partial{\bf A}^{\rm T}}{\partial x_{s}}
\end{array}\right] \in \mathds{R}^{r{\times}q{\times}s} $.}
\entry{${\bf A}_{\bf x}{\bf b}$}{$= {\bf A}_{\bf x}\otimes{\bf b} =
\left[\begin{array}{c c c c c}
\displaystyle\frac{\partial{\bf A}}{\partial x_1} {\bf b} & \ldots &
\displaystyle\frac{\partial{\bf A}}{\partial x_i} {\bf b} & \ldots &
\displaystyle\frac{\partial{\bf A}}{\partial x_{s}} {\bf b}
\end{array}\right] \in \mathds{R}^{q{\times}s}$, where ${\bf b} \in \mathds{R}^{r}$ is a vector.}
\entry{${\bf A}_{\bf x}{\bf B}$}{$= {\bf A}_{\bf x}\otimes{\bf B} =
\left[\begin{array}{c c c c c}
\displaystyle\frac{\partial{\bf A}}{\partial x_1} {\bf B} & \ldots &
\displaystyle\frac{\partial{\bf A}}{\partial x_i} {\bf B} & \ldots &
\displaystyle\frac{\partial{\bf A}}{\partial x_{s}} {\bf B}
\end{array}\right] \in \mathds{R}^{q{\times}t{\times}s}$, where ${\bf B} \in \mathds{R}^{r{\times}t}$ is a matrix.}
\entry{${\bf C} {\bf A}_{\bf x}{\bf B}$}{$={\bf C} \otimes {\bf A}_{\bf x}{\bf B} =
\left[\begin{array}{c c c c c}
{\bf C} \displaystyle\frac{\partial{\bf A}}{\partial x_1} {\bf B} & \ldots &
{\bf C} \displaystyle\frac{\partial{\bf A}}{\partial x_i} {\bf B} & \ldots &
{\bf C} \displaystyle\frac{\partial{\bf A}}{\partial x_{s}} {\bf B}
\end{array}\right] \in \mathds{R}^{r{\times}t{\times}s}$, where ${\bf C} \in \mathds{R}^{r{\times}q}$ is a matrix.}
\end{nomenclature}

\section{Introduction}
Multibody dynamics has become an essential tool for mechanical systems analysis. The progress made during the last decades lead to the development of advanced multibody dynamics techniques and complex models that can account for phenomena otherwise difficult to consider and not feasible to achieve with analytical models. One important opportunity to expand the state-of-the-art research in multibody models is the design optimization of multibody systems with respect to design parameters. Sensitivity analysis of the dynamic response of multibody systems is essential for gradient-based optimization.

Numerical sensitivities, when needed, are often calculated by means of finite differences. However, most of the time, the objective function is not only related to the design parameters, but also related to the state variables of the equation of motion. Due to this reason, in order to obtain the numerical sensitivities, the equation of motion must be solved repeatedly. Thus, to calculate numerical sensitivities is computationally expensive. Moreover, in many cases, the results suffer from low accuracy due to computer round-off errors.

Due to the shortcomings of numerical sensitivities, development of analytical approaches to perform sensitivity analysis becomes essential. There are two well-known sensitivity approaches: the direct sensitivity approach and the adjoint sensitivity approach. Haug and Arora, 1978, first presented the adjoint sensitivity approach \cite{Haug1978}. In a later study,  the sensitivity analysis of dynamic mechanical systems was presented by Haug, Wehage, and Mani, 1984 \cite{Haug1984}. The direct sensitivity approach was presented in the same year by Krishnaswami and Bhatti \cite{Bhatti1984}. Methodologies based on these two sensitivity approaches, for various multibody formulations, have then been developed. For example, the direct sensitivity approaches using index-3 and 1ndex-1 differential algebraic equations (DAEs) formulations were developed by Haug in 1987 \cite{Haug1987} and Chang in 1985 \cite{chang1985}; the direct sensitivity approaches using penalty and augmented Lagrangian formulations was developed by Pagalday in 1997 \cite{Pagalday1997}; the ajoint sensitivity methods using index-3 and index-1 DAEs formulations were developed by Haug in 1981 \cite{haug1981design}, Haug in 1987 \cite{Haug1987}, and Bestle in 1992 \cite{Bestle1992}. For more sensitivity studies, the reader is referred to \cite{Bestle1992a,Dias1997,Feehery1997,Anderson2002,Anderson2004,Ding2007,Schaffer2006,Neto2009,Bhalerao2010,Banerjee2013}. 
These methods have some drawbacks that prevent them from easily computing sensitivities for large and complex multibody systems with respect to a large number of design parameters. For instance, the direct sensitivity approach works well when the number of parameters is small, but it becomes computationally expensive when the number of parameters is large. On the other hand, the ajoint sensitivity methods using index-3 and index-1 DAEs formulations are not computationally efficient because of the numerical difficulties to solve DAEs \cite{Brenan1989,Ascher1998}.

Thus, the main task of the study presented in this paper is to overcome these drawbacks and to create a new approach in order to efficiently perform sensitivity analysis and optimization for large and complex systems with respect to a large number of parameters.

Unlike the direct sensitivity approach, the adjoint sensitivity approach works well when the number of parameters is large. On the other hand, solving ordinary differential equations (ODEs) is computationally easier than solving DAEs. Thus, the adjoint sensitivity approaches using ODEs formulations become popular. Dopico and Zhu, 2014, first developed the adjoint sensitivity approach using ODEs formulation \cite{dopicodirect}. In that paper, direct and adjoint sensitivity approaches are developed for the state-space formulation based on the projection Matrix R \cite{GarciadeJalon1994}, or Maggi’s equations. For Maggi’s formulation, since the dynamic equations are transformed from dependent to independent coordinates at each time step, the approach is not stable when the multibody system goes through a singular or bifurcation position \cite{jalon1994kinematic}.

In this paper, the penalty formulation is used to compute sensitivities. The penalty formulation is an ODEs formulation with dependent coordinates, which was introduced in \cite{Bayo1988,GarciadeJalon1994}. Comparing this formulation with the DAEs formulations and Maggi's formulation, the penalty formulation is more stable, and it doesn't fail around kinematic singularity; it also allows redundant constraints. Moreover, it is more computationally efficient to be solved than DAEs formulations \cite{jalon1994kinematic}. In addition, unlike Maggi's formulation, the penalty formulation doesn't need to restart the numerical integrator for each time step. The shortcoming of the penalty formulation is  that it requires an arbitrary value for its penalty factor and for two other coefficients. There is no rigorous method of determining acceptable values for these terms. This penalty factor is typically chosen based on the researcher's experience with this formulation. For example, $10^9$ is chosen as the value of the penalty factor in this paper and it works perfectly well. The other two coefficients are usually chosen such as to have critical damping in the vibrations associated to the constraints.

The new theory developed is demonstrated on one academic case study,  a five-bar mechanism, and on one real-life system,  a 14-\textit{DOF} vehicle model. The five-bar mechanism is used to illustrate the sensitivity approach derived in this paper. The full vehicle model is used to demonstrate the capability of the new approach developed to perform sensitivity analysis and gradient-based optimization for large and complex multibody systems with respect to multiple design parameters.

Finally, using the outputs of the sensitivity analysis, a gradient-based optimization package (L-BFGS-B) \cite{lbfgsb} is employed to perform the dynamical optimization of the full vehicle system with respect to 6 design parameters.


\section{Design optimization of mechanical systems}

The design optimization of a mechanical system usually concerns a set of design parameters $\bm \rho \in \mathds{R}^{p}$; these parameters are related to the geometry, materials, or other characteristics that must be specified by the design engineer. The optimization theory can considerably help the engineer make such decisions.

The objective of the optimization is to find the values of the selected design parameters that produce the best performance/behavior of the system, under the given constraints. The behavior of the system is represented mathematically by a cost or objective function  $\psi = \psi \left({\bm \rho}\right)$, which is minimized by the optimal value of the parameters.

In cases where the optimization is based on the dynamical behavior of the system under given inputs and initial conditions, the objective function often depends directly on the states of the system in the form $\psi = \psi \left({\bf y}\right)$. The system states depend on the parameters ${\bf y} = {\bf y} \left( {\bm \rho} \right)$ through the dynamics of the system.

It is also quite usual that the vector of design variables cannot have any values while in the same time it is subjected to some design constraints. The design constraints should be equality or inequality relations, e.g., ${\bm \Psi} \left({\bm \rho}\right) = 0$.

Many advanced numerical optimization methods require the gradient of the objective function/constraints with respect to the parameters. In this paper, based on the outputs of the sensitivity analysis, a gradient-based optimization package (L-BFGS-B) is applied to perform the dynamical optimization. Subsequent sections present the approach developed to perform this sensitivity analysis.

\section{Description of the multibody formulation}
\label{sc:ALform}
The equations of motion (EOM) in the penalty formulation \cite{Bayo1988} have the following expression
\begin{equation}
\label{eq:penalty}
{\bf M} \ddot{\bf q} + {\bm \Phi}_{\bf q}^{\rm T}{\alpha}\left(\ddot{\bm \Phi}+
2 \xi \omega \dot{\bm \Phi}+\omega^2 {\bm \Phi}\right) = {\bf Q} \, ,
\end{equation}
where $\alpha$ is the penalty factor, $\xi$ and $\omega$ are coefficients of the method, and the rest of the terms are described in the nomenclature.

Equations \eqref{eq:penalty} constitute an ordinary differential equation (ODE) that replaces the constraints of the original index-3 DAEs system by a penalty term that makes possible to estimate the Lagrange multipliers associated to the constraint reactions by the following formula:
\begin{equation}
\label{eq:lambda}
{\bm \lambda}^{*}={\alpha}\left(\ddot{\bm \Phi}+
2 \xi \omega \dot{\bm \Phi}+\omega^2 {\bm \Phi}\right)\,.
\end{equation}

Expanding equation \eqref{eq:penalty}, one obtains the following second order ODE system with accelerations as unknowns:
\begin{eqnarray}
\label{eq:penalty_compact}
\label{eq:penalty_compacta}
&&\bar{\bf M} \left({\bf q},{\bm \rho}\right) \ddot{\bf q} = 
\bar{\bf Q} \left({\bf q},\dot{\bf q},t,{\bm \rho}\right) \\
\label{eq_barM_penalty}
&&\bar{\bf M} = {\bf M}+{\bm \Phi}_{\bf q}^{\rm T}{\alpha}{\bm \Phi}_{\bf q} \\
\label{eq_barQ_penalty}
&&\bar{\bf Q} = {\bf Q}
-{\bm \Phi}_{\bf q}^{\rm T}{\alpha}\left(\dot{\bm \Phi}_{\bf q}\dot{\bf q}+
\dot{\bm \Phi}_{t}+2 \xi \omega \dot{\bm \Phi}+\omega^2 {\bm \Phi}\right)
\end{eqnarray}
where the following kinematic identities hold
\begin{eqnarray}
\label{eq:dotPhi}
&&\dot{\bm \Phi}={\bm \Phi}_{\bf q}\dot{\bf q}+{\bm \Phi}_{t} \,, \\
\label{eq:ddotPhi}
&&\ddot{\bm \Phi}={\bm \Phi}_{\bf q}\ddot{\bf q}
+{\dot{\bm \Phi}}_{\bf q}\dot{\bf q} + {\dot{\bm \Phi}}_{t} \,.
\end{eqnarray}

Note that the EOM \eqref{eq:penalty_compact} depend on some design parameters $\bm \rho \in \mathds{R}^{p}$ (typically masses, lengths, or other parameters related to forces chosen by the engineer). Therefore ${\bf q}= {\bf q}\left(t,{\bm \rho}\right)$, $\dot{\bf q}= \dot{\bf q}\left(t,{\bm \rho}\right)$, and $\ddot{\bf q}= \ddot{\bf q}\left(t,{\bm \rho}\right)$.

\section{Adjoint sensitivity approach}

The adjoint approach seeks to obtain the sensitivity of a cost function, $\psi$, with respect to the set of parameters $\bm \rho$. For practical applications, very general cost functions depend not only on positions and velocities, but also on accelerations and reaction forces:
\begin{equation}
\label{eq_cost_function}
{\psi} = {w}\left({\bf q}_{F},\dot{\bf q}_{F},\ddot{\bf q}_{F},{\bm \rho},{\bm \lambda}_{F}^{*} \right) + 
\int_{t_0}^{t_F} {g}\left({\bf q},\dot{\bf q},\ddot{\bf q},{\bm \rho},{\bm \lambda}^{*}\right){\rm dt} \, .
\end{equation}
The system \eqref{eq:penalty_compact} can be transformed into a first order system by simply defining a new set of variables by the relation $\dot{\bf q} = {\bf v}$,
\begin{eqnarray}
\left[ \begin{array}{c c}
{\bf I} & {\bf 0} \\
{\bf 0} & \bar{\bf M}
\end{array} \right]  
\left[ \begin{array}{c}
 \dot{\bf q} \\
 \dot{\bf v}
\end{array} \right] = &
\left[ \begin{array}{c}
 {\bf v} \\
 \bar{\bf Q}
\end{array} \right]  \quad \Leftrightarrow \quad
\label{eq_1st_csemiexplicit_ODE}
\hat{\bf M} \left({\bf y},{\bm \rho}\right) \dot{\bf y} =   
\hat{\bf Q} \left(t,{\bf y},{\bm \rho} \right).~~
\end{eqnarray}

In \eqref{eq_1st_csemiexplicit_ODE}, the new state vector is ${\bf y} = \left[ \begin{array}{c c}  {\bf q}^{\rm T} & {\bf v}^{\rm T}  \end{array} \right]^{\rm T}$.
Taking the inverse of the leading matrix, the system \eqref{eq_1st_csemiexplicit_ODE} can be expressed as a first order explicit ODE
\begin{equation}
\label{eq:1st_explicit_ODE}
\dot{\bf y} =  \hat{\bf M}^{\rm -1} \left({\bf y},{\bm \rho}\right)
\hat{\bf Q} \left(t,{\bf y},{\bm \rho} \right) = 
{\bf f}\left(t,{\bf y},{\bm \rho}\right).
\end{equation}
Similarly, the objective function \eqref{eq_cost_function} can be expressed as a function of the first order states
\begin{equation}
\label{eq_cost_function2}
{\psi} ={w}\left({\bf y}_{F},\dot{\bf y}_{F},{\bm \rho}_{F},{\bm \lambda}_{F}^{*} \right) + 
\int_{t_0}^{t_F} {g}\left({\bf y},\dot{\bf y},{\bm \rho},{\bm \lambda}^{*}\right){\rm dt} \, .
\end{equation}

Following \cite{Cao2003}, we consider the following Lagrangian, given by the cost function subject to the EOM constraints
\begin{equation}
\label{eq_Lagrangian}
{\mathcal L} \left({\bm \rho}\right)= {\psi} 
-\int_{t_0}^{t_F} {\bm \mu}^{\rm T} \left(\dot{\bf y} - 
{\bf f}\left(t,{\bf y},{\bm \rho}\right) \right){\rm dt},
\end{equation}
where $\bm \mu$ is the vector of Lagrange multipliers or adjoint variables. Applying variational calculus
\begin{eqnarray}
\label{eq_varLagrangian}
&&{\delta {\mathcal L}} =  \delta{\psi}
- \int_{t_0}^{t_F} {\delta{\bm \mu}^{\rm T} \left(\dot{\bf y} - 
{\bf f}\left(t,{\bf y},{\bm \rho}\right)\right) {\rm dt}} \nonumber \\
&&- \int_{t_0}^{t_F} {{\bm \mu}^{\rm T} \left(\delta \dot{\bf y} - 
{\bf f}_{\bf y} \delta{\bf y} -
{\bf f}_{\bm \rho} \delta{\bm \rho} \right) {\rm dt}}
\end{eqnarray}
The central term vanishes if the EOM are fulfilled at each time step. 

The variation of the cost function is
\begin{eqnarray}
\label{eq:delta_psi}
&&\delta{\psi}= \left( {w}_{\bf y} \delta{\bf y} +
{w}_{\dot{\bf y}} \delta{\dot{\bf y}}
+{w}_{\bm \rho} \delta{\bm \rho} 
+{w}_{\bm \lambda^{*}} \delta{\bm \lambda}^* \right)_{F}+ \nonumber \\
&&\int_{t_0}^{t_F} {\left( {g}_{\bf y} \delta{\bf y} 
+ {g}_{\dot{\bf y}} \delta{\dot{\bf y}}+
{g}_{\bm \rho} \delta{\bm \rho} +
{g}_{{\bm \lambda}^{*}} \delta{\bm \lambda}^* \right){\rm dt}}.
\end{eqnarray}
From Eqn.~\eqref{eq:lambda}
\begin{subequations}
\begin{equation}
\label{eq:delta_lambda}
\delta{\bm \lambda}^{*} = {\alpha} \left( 
{\delta{\ddot{\bm \Phi}}}+2 {\xi} {\omega}{\delta{\dot{\bm \Phi}}}+
{\omega}^2 {\delta{\bm \Phi}} \right)
\, ,
\end{equation}
where
\begin{eqnarray}
\label{eq:deltafis}
&&{\delta{\ddot{\bm \Phi}}} = {\bm \Phi}_{\bf q} {\delta{\ddot{\bf q}}}
+\left({\bm \Phi}_{\bf qq}\dot{\bf q} + \dot{\bm \Phi}_{\bf q} + {\bm \Phi}_{t{\bf q}} \right) 
{\delta{\dot{\bf q}}} \nonumber \\
&& +\left({\bm \Phi}_{\bf qq}\ddot{\bf q} 
+\left( {\dot{\bm \Phi}}_{\bf q}\right)_{\bf q} {\dot{\bf q}}+
\left({\dot{\bm \Phi}}_{t}\right)_{\bf q} \right) {\delta{\bf q}} \nonumber \\ && +\left( {\bm \Phi}_{{\bf q}{\bm \rho}}{\ddot{\bf q}}+
\left(\dot{\bm \Phi}_{\bf q}\right)_{\bm \rho}{\dot{\bf q}} +\left({\dot{\bm \Phi}}_{t}\right)_{\bm \rho} \right){\delta{\bm \rho}} \\
\label{eq:deltafip}
&&{\delta{\dot{\bm \Phi}}}={\bm \Phi}_{\bf q}{\delta{\dot{\bf q}}} 
\!+\!\left({\bm \Phi}_{\bf qq}\dot{\bf q}\!+\!{\bm \Phi}_{t{\bf q}}\right) {\delta{\bf q}}
\!+\!\left({\bm \Phi}_{{\bf q}{\bm \rho}}\dot{\bf q}\!+\!{\bm \Phi}_{t{\bm \rho}}\right) {\delta{\bm \rho}} \quad \\
\label{eq:deltafi}
&&{\delta{\bm \Phi}} = {\bm \Phi}_{\bf q}{\delta{\bf q}} +{\bm \Phi}_{\bm \rho} {\delta{\bm \rho}} 
\end{eqnarray}
\end{subequations}

Grouping together the terms associated to ${\delta{\ddot{\bf q}}}$, ${\delta{\dot{\bf q}}}$, ${\delta{\bf q}}$, ${\delta{\bm \rho}}$ and taking into account that ${\bf y} = \left[ \begin{array}{c c}  {\bf q}^{\rm T} & {\bf v}^{\rm T}  \end{array} \right]^{\rm T}$, Eqn.~\eqref{eq:delta_lambda} becomes

\begin{equation}
\label{eq:delta_lambda2}
\delta{\bm \lambda}^{*} = 
{\bm \lambda}_{\dot{\bf y}}^{*} {\delta{\dot{\bf y}}}+ 
{\bm \lambda}_{\bf y}^{*} {\delta{\bf y}}+ 
{\bm \lambda}_{\bm \rho}^{*} {\delta{\bm \rho}} \, ,
\end{equation}

Identifying the common terms in \eqref{eq:delta_lambda} and \eqref{eq:delta_lambda2} and using the identity ${\bf v} = {\dot{\bf q}}$ one obtains
\begin{eqnarray}
\label{eq:lambda_y}
&&{\bm \lambda}_{\bf y} = \left[\begin{array}{c c}
{\bm \lambda}_{\bf q}^{*} & {\bm \lambda}_{\bf v}^{*}
\end{array}\right] \\
\label{eq:lambda_yp}
&&{\bm \lambda}_{\dot{\bf y}} = \left[\begin{array}{c c}
{\bf 0} & {\bm \lambda}_{\dot{\bf v}}^{*}
\end{array}\right] \\
\label{eq:lambda_vp}
&&{\bm \lambda}_{\dot{\bf v}}^{*} = {\alpha} {\bm \Phi}_{\bf q} \\
\label{eq:lambda_v}
&&{\bm \lambda}_{\bf v}^{*} = {\alpha} 
\left[{\bm \Phi}_{\bf qq}{\bf v} + \dot{\bm \Phi}_{\bf q} + {\bm \Phi}_{t{\bf q}}
+2 {\xi} {\omega} {\bm \Phi}_{\bf q} \right] \\
\label{eq:lambda_q}
&&{\bm \lambda}_{\bf q}^{*} = {\alpha} 
\left[ {\bm \Phi}_{\bf qq}{\dot{\bf v}} 
+\left( {\dot{\bm \Phi}}_{\bf q}\right)_{\bf q} {\bf v}+
\left({\dot{\bm \Phi}}_{t}\right)_{\bf q} \right. \nonumber \\
&& \left. +2 {\xi} {\omega} \left({\bm \Phi}_{\bf qq}{\bf v} + {\bm \Phi}_{t{\bf q}}\right) 
+{\omega}^2 {\bm \Phi}_{\bf q} \right] \\
\label{eq:lambda_ro}
&&{\bm \lambda}_{\bm \rho}^{*} = {\alpha} 
\left[ {\bm \Phi}_{{\bf q}{\bm \rho}}{\dot{\bf v}}+
\left(\dot{\bm \Phi}_{\bf q}\right)_{\bm \rho}{\bf v} +\left({\dot{\bm \Phi}}_{t}\right)_{\bm \rho} \right. \nonumber \\
&&\left. +2 {\xi} {\omega} \left({\bm \Phi}_{{\bf q}{\bm \rho}}{\bf v} +{\bm \Phi}_{t{\bm \rho}}\right)
+{\omega}^2 {\bm \Phi}_{\bm \rho}
\right]
\end{eqnarray}
Replacing \eqref{eq:delta_lambda2} in \eqref{eq:delta_psi}
\begin{eqnarray}
\label{eq:delta_psiB}
&&\delta{\psi}= \left[ 
\left( {w}_{\bf y} + {w}_{\bm \lambda^{*}} {\bm \lambda}_{\bf y}^{*} \right) \delta{\bf y} 
+\left( {w}_{\dot{\bf y}} + {w}_{\bm \lambda^{*}} {\bm \lambda}_{\dot{\bf y}}^{*} \right) \delta{\dot{\bf y}} \right. \nonumber \\
&&\left. +\left( {w}_{\bm \rho} + {w}_{\bm \lambda^{*}} {\bm \lambda}_{\bm \rho}^{*} \right) \delta{\bm \rho} \right]_{F}
+\int_{t_0}^{t_F} \left[ 
\left( {g}_{\bf y} + {g}_{{\bm \lambda}^{*}} {\bm \lambda}_{\bf y}^{*}\right) \delta{\bf y} \right. \nonumber \\
&&\left. +\left( {g}_{\dot{\bf y}} + {g}_{{\bm \lambda}^{*}} {\bm \lambda}_{\dot{\bf y}}^{*} \right)\delta{\dot{\bf y}}
+\left( {g}_{\bm \rho} + {g}_{{\bm \lambda}^{*}} {\bm \lambda}_{\bm \rho}^{*} \right)\delta{\bm \rho}
\right]{\rm dt}
\end{eqnarray}

For convenience, $\delta{\dot{\bf y}}$ in \eqref{eq:delta_psiB} can be expressed as a function of $\delta{\bf y}$. Differentiating Eqn.~\eqref{eq:1st_explicit_ODE}
\begin{equation}
\label{eq:deltayp}
\delta{\dot{\bf y}} = {\bf f}_{\bf y} \delta{\bf y} + {\bf f}_{\bm \rho} \delta{\bm \rho}
\end{equation}
and replacing Eqn.~\eqref{eq:deltayp} in \eqref{eq:delta_psiB} leads to
\begin{eqnarray}
\label{eq:delta_psiC}
&&\delta{\psi}= \left[ \left( {w}_{\bf y} 
+ {w}_{\bm \lambda^{*}} {\bm \lambda}_{\bf y}^{*}
+ \left( {w}_{\dot{\bf y}} + {w}_{\bm \lambda^{*}} {\bm \lambda}_{\dot{\bf y}}^{*} \right) 
{\bf f}_{\bf y} \right) \delta{\bf y} \right. \nonumber \\
&&\left. + \left( {w}_{\bm \rho} 
+ {w}_{\bm \lambda^{*}} {\bm \lambda}_{\bm \rho}^{*}
+ \left( {w}_{\dot{\bf y}} + {w}_{\bm \lambda^{*}} {\bm \lambda}_{\dot{\bf y}}^{*} \right) {\bf f}_{\bm \rho}
\right) \delta{\bm \rho} \right]_{F} \nonumber \\
&&+\int_{t_0}^{t_F} \left[ 
\left( {g}_{\bf y} + {g}_{{\bm \lambda}^{*}} {\bm \lambda}_{\bf y}^{*}
+ \left( {g}_{\dot{\bf y}} 
+ {g}_{{\bm \lambda}^{*}} {\bm \lambda}_{\dot{\bf y}}^{*} \right) {\bf f}_{\bf y} \right) \delta{\bf y} \right. \nonumber \\
&&\left. +\left( {g}_{\bm \rho} + {g}_{{\bm \lambda}^{*}} {\bm \lambda}_{\bm \rho}^{*}
+ \left( {g}_{\dot{\bf y}} 
+ {g}_{{\bm \lambda}^{*}} {\bm \lambda}_{\dot{\bf y}}^{*} \right) {\bf f}_{\bm \rho} \right)\delta{\bm \rho}
\right]{\rm dt}.
\end{eqnarray}

The variation of the full Lagrangian \eqref{eq_varLagrangian} can be obtained by replacing \eqref{eq:delta_psiC} in \eqref{eq_varLagrangian}
\begin{eqnarray}
\label{eq:varLagrangianB}
&&{\delta {\mathcal L}} =  
\left[ \left( {w}_{\bf y} 
+ {w}_{\bm \lambda^{*}} {\bm \lambda}_{\bf y}^{*}
+ \left( {w}_{\dot{\bf y}} + {w}_{\bm \lambda^{*}} {\bm \lambda}_{\dot{\bf y}}^{*} \right) 
{\bf f}_{\bf y} \right) \delta{\bf y} \right. \nonumber \\
&&\left. + \left( {w}_{\bm \rho} 
+ {w}_{\bm \lambda^{*}} {\bm \lambda}_{\bm \rho}^{*}
+ \left( {w}_{\dot{\bf y}} + {w}_{\bm \lambda^{*}} {\bm \lambda}_{\dot{\bf y}}^{*} \right) {\bf f}_{\bm \rho}
\right) \delta{\bm \rho} \right]_{F} \nonumber \\
&&+\int_{t_0}^{t_F} \left[ 
\left( {g}_{\bf y} + {g}_{{\bm \lambda}^{*}} {\bm \lambda}_{\bf y}^{*}
+ \left( {\bm \mu}^{\rm T} + {g}_{\dot{\bf y}} 
+ {g}_{{\bm \lambda}^{*}} {\bm \lambda}_{\dot{\bf y}}^{*} \right) {\bf f}_{\bf y} \right) \delta{\bf y} \right. \nonumber \\
&&\left. \!+\!\left({g}_{\bm \rho} \!+\! {g}_{{\bm \lambda}^{*}} {\bm \lambda}_{\bm \rho}^{*}
\!+\! \left( {\bm \mu}^{\rm T} \!+\! {g}_{\dot{\bf y}} 
\!+\! {g}_{{\bm \lambda}^{*}} {\bm \lambda}_{\dot{\bf y}}^{*} \right) {\bf f}_{\bm \rho} \right)\delta{\bm \rho}
\!-\! {\bm \mu}^{\rm T} \delta{\dot{\bf y}} \right]{\rm dt}
\end{eqnarray}

In Eqn.~\eqref{eq:varLagrangianB}, the variation of the parameters $\delta{\bm \rho}$ is known, and variations $\delta{\bf y}$ and $\delta{\dot{\bf y}}$ could be calculated by solving the linearized form of the EOM \eqref{eq:1st_explicit_ODE}, but this is computationally expensive. Instead of calculating them, the idea is to cancel these variations. Integrating by parts the integral terms involving $\delta{\dot{\bf y}}$ the variation can be removed from the integral

\begin{equation}
\label{eq_bypartsInt}
\int_{t_0}^{t_F} - {\bm \mu}^{\rm T} \delta{\dot{\bf y}} {\rm dt}=
\left. - {\bm \mu}^{\rm T} \delta{\bf y} \right|_{t_0}^{t_F}
+\int_{t_0}^{t_F} {\dot{\bm \mu}}^{\rm T} \delta{\bf y} {\rm dt}.
\end{equation}

Therefore
\begin{eqnarray}
\label{eq:varLagrangianC}
&&{\delta {\mathcal L}} =  
\left[ \left( {w}_{\bf y} 
+ {w}_{\bm \lambda^{*}} {\bm \lambda}_{\bf y}^{*}
+ \left( {w}_{\dot{\bf y}} + {w}_{\bm \lambda^{*}} {\bm \lambda}_{\dot{\bf y}}^{*} \right) 
{\bf f}_{\bf y} - {\bm \mu}^{\rm T} \right) \delta{\bf y} \right. \nonumber \\
&&\left. + \left( {w}_{\bm \rho} 
+ {w}_{\bm \lambda^{*}} {\bm \lambda}_{\bm \rho}^{*}
+ \left( {w}_{\dot{\bf y}} + {w}_{\bm \lambda^{*}} {\bm \lambda}_{\dot{\bf y}}^{*} \right) {\bf f}_{\bm \rho}
\right) \delta{\bm \rho} \right]_{F}
+\left[ {\bm \mu}^{\rm T} \delta{\bf y} \right]_{0} \nonumber \\
&&+\int_{t_0}^{t_F} \left[ 
\left( {\dot{\bm \mu}}^{\rm T} + {g}_{\bf y} + {g}_{{\bm \lambda}^{*}} {\bm \lambda}_{\bf y}^{*}
+ \left( {\bm \mu}^{\rm T} + {g}_{\dot{\bf y}} 
+ {g}_{{\bm \lambda}^{*}}{\bm \lambda}_{\dot{\bf y}}^{*} \right) {\bf f}_{\bf y} \right) \delta{\bf y} \right. \nonumber \\
&&\left. +\left( {g}_{\bm \rho} + {g}_{{\bm \lambda}^{*}} {\bm \lambda}_{\bm \rho}^{*}
+ \left( {\bm \mu}^{\rm T} + {g}_{\dot{\bf y}} 
+ {g}_{{\bm \lambda}^{*}} {\bm \lambda}_{\dot{\bf y}}^{*} \right) {\bf f}_{\bm \rho} \right)\delta{\bm \rho}\right]{\rm dt}.
\end{eqnarray}

In Eqn. \eqref{eq:varLagrangianC} it is possible to cancel $\delta{\bf y}$ by choosing ${\bm \mu}$ to be the solution of following {\em adjoint ODE system}
\begin{eqnarray}
\label{eq:adjointODE}
&&{\dot{\bm \mu}} = - {\bf f}_{\bf y}^{\rm T} \left( {\bm \mu} + {g}_{\dot{\bf y}}^{\rm T}
+ {\bm \lambda}_{\dot{\bf y}}^{*{\rm T}} {g}_{{\bm \lambda}^{*}}^{\rm T} \right)
-{g}_{\bf y}^{\rm T} - {\bm \lambda}_{\bf y}^{*{\rm T}} {g}_{{\bm \lambda}^{*}}^{\rm T}, \\
\label{eq:adjointODEb}
&&{\bm \mu}_{F} =
\left[ {w}_{\bf y}^{\rm T} + {\bm \lambda}_{\bf y}^{*{\rm T}} {w}_{\bm \lambda^{*}}^{\rm T} 
+ {\bf f}_{\bf y}^{\rm T} \left( {w}_{\dot{\bf y}} + {w}_{\bm \lambda^{*}} {\bm \lambda}_{\dot{\bf y}}^{*} \right)^{\rm T}\right]_F.
\end{eqnarray}
The adjoint system \eqref{eq:adjointODE} is a first order linear ODE in $\bm \mu$. Since the initial conditions \eqref{eq:adjointODEb} are given at the final time $t_F$, it has to be integrated backward in time from $t_F$ to $t_0$ as an initial value problem.

Finally, from Eqn. \eqref{eq:varLagrangianC} the gradient of the cost function with respect to parameters can be obtained as
\begin{eqnarray}
\label{eq_gradient}
&&{\nabla_{\bm \rho} \psi} = 
\left[{w}_{\bm \rho}^{\rm T} 
+ {\bm \lambda}_{\bm \rho}^{*{\rm T}} {w}_{\bm \lambda^{*}}^{\rm T}
+ {\bf f}_{\bm \rho}^{\rm T}
\left( {w}_{\dot{\bf y}} + {w}_{\bm \lambda^{*}} {\bm \lambda}_{\dot{\bf y}}^{*} \right)^{\rm T} \right]_{F}
+\left[ \frac{\partial{\bf y}^{\rm T}}{\partial {\bm \rho}} {\bm \mu} \right]_{0} \nonumber \\
&&+\int_{t_0}^{t_F} \left[
{\bf f}_{\bm \rho}^{\rm T} \left( {\bm \mu} + {g}_{\dot{\bf y}}^{\rm T} 
+ {\bm \lambda}_{\dot{\bf y}}^{*{\rm T}} {g}_{{\bm \lambda}^{*}}^{\rm T} \right) 
+ {g}_{\bm \rho}^{\rm T} + {\bm \lambda}_{\bm \rho}^{*{\rm T}} {g}_{{\bm \lambda}^{*}}^{\rm T}
\right]{\rm dt} \, ,
\end{eqnarray}
where the identity $\delta \psi = \delta L $ was used. This holds if the EOM are satisfied, as can be seen from Eqn.~\eqref{eq_Lagrangian}.

In Eqns.~\eqref{eq_gradient} and \eqref{eq:adjointODE} the derivatives of function $g$ are known, since the objective function has a known expression. The derivatives of $\bf f$ are obtained using \eqref{eq_1st_csemiexplicit_ODE} as
\begin{subequations}
\begin{gather}
\label{eq_dfdy}
\hat{\bf M} \frac{\partial{\bf f}}{\partial{\bf y}}  + 
{\hat{\bf M}}_{\bf y} {\bf f} =  \frac{\partial \hat{\bf Q}}{\partial{\bf y}} ~~\Rightarrow~~ 
{\bf f}_{\bf y} = \hat{\bf M}^{\rm -1} \left(
{\hat{\bf Q}}_{\bf y} - {\hat{\bf M}}_{\bf y} {\bf f} \right), \\
\label{eq_dfdro}
\hat{\bf M} \frac{\partial{\bf f}}{\partial{\bm \rho}} + 
{\hat{\bf M}}_{\bm \rho} {\bf f} =  \frac{\partial \hat{\bf Q}}{\partial{\bm \rho}} ~~\Rightarrow~~ 
{\bf f}_{\bm \rho} = \hat{\bf M}^{\rm -1} \left(
{\hat{\bf Q}}_{\bm \rho} - {\hat{\bf M}}_{\bm \rho} {\bf f} \right).
\end{gather}
\end{subequations}
The derivatives ${\bf f}_{\bf y}$ and ${\bf f}_{\bm \rho}$ can be calculated in block form as
\begin{eqnarray}
\label{eq:block_dfdy}
&&{\bf f}_{\bf y} = 
\left[ \begin{array}{c c}
{\bf I} & {\bf 0} \\
{\bf 0} & \bar{\bf M}^{\rm -1}
\end{array} \right]  \left( 
\left[ \begin{array}{c c}
{\bf 0} & {\bf I} \\
-\bar{\bf K} & -\bar{\bf C}
\end{array} \right] -
\left[ \begin{array}{c c}
{\bf 0} & {\bf 0} \\
{\bar{\bf M}}_{\bf q} {\dot{\bf v}} & {\bf 0} 
\end{array} \right]
\right) = \nonumber \\
&&\left[ \begin{array}{c c}
{\bf 0} & {\bf I} \\
-\bar{\bf M}^{\rm -1}\left(\bar{\bf K}+
\bar{\bf M}_{\bf q} \dot{\bf v}\right) & 
-\bar{\bf M}^{\rm -1}\bar{\bf C}
\end{array} \right] \,, \\
\label{eq:block_dfdro}
&&{\bf f}_{\bm \rho} = 
\left[ \begin{array}{c c}
{\bf I} & {\bf 0} \\
{\bf 0} & \bar{\bf M}
\end{array} \right]^{\rm -1}
\left(\left[ \begin{array}{c}
{\bf 0} \\
{\bar{\bf Q}}_{\bm \rho}
\end{array} \right]
-\left[ \begin{array}{c}
{\bf 0} \\
{\bar{\bf M}}_{\bm \rho} {\dot{\bf v}}
\end{array} \right] \right) = \nonumber \\
&&\left[ \begin{array}{c}
{\bf 0} \\
\bar{\bf M}^{\rm -1}\left(\bar{\bf Q}_{\bm \rho} - \bar{\bf M}_{\bm \rho} \dot{\bf v} \right)
\end{array} \right]\,.
\end{eqnarray}

In Eqns.~\eqref{eq:block_dfdy} and \eqref{eq:block_dfdro} the terms $\bar{\bf K}$, $\bar{\bf C}$, $\bar{\bf Q}_{\bm \rho}$, $\bar{\bf M}_{\bf q} \ddot{\bf q}$, and $\bar{\bf M}_{\bm \rho} \ddot{\bf q}$ are given by the following expressions:
\begin{eqnarray}
\label{eq:barK}
&&\bar{\bf K} = -\frac{\partial\bar{\bf Q}}{\partial{\bf q}} = {\bf K} + 
{\bm \Phi}_{\bf qq}^{\rm T}{\alpha}\left(\dot{\bm \Phi}_{\bf q}\dot{\bf q}+
\dot{\bm \Phi}_{t}+2 \xi \omega \dot{\bm \Phi}+\omega^2 {\bm \Phi}\right)+\nonumber \\
&&{\bm \Phi}_{\bf q}^{\rm T}{\alpha}\left(\left(\dot{\bm \Phi}_{\bf q}\dot{\bf q}\right)_{\bf q}\!+\!
\left(\dot{\bm \Phi}_{t}\right)_{\bf q}\!+\!2 \xi \omega 
\left({\bm \Phi}_{\bf qq}\dot{\bf q}\!+\!{\bm \Phi}_{t{\bf q}}\right)\!+\! 
\omega^2 {\bm \Phi}_{\bf q}\right), \\
\label{eq:barC}
&&\bar{\bf C}\!=\!-\frac{\partial\bar{\bf Q}}{\partial\dot{\bf q}}\!=\!{\bf C}\!+\! 
{\bm \Phi}_{\bf q}^{\rm T}{\alpha}\left({\bm \Phi}_{\bf qq}\dot{\bf q}\!+\!\dot{\bm \Phi}_{\bf q}\!+\!
{\bm \Phi}_{tq}+2 \xi \omega {\bm \Phi}_{\bf q}\right), \\
\label{eq_dbarQdro}
&&\bar{\bf Q}_{\bm \rho}=\frac{\partial\bar{\bf Q}}{\partial{\bm \rho}} = {\bf Q}_{\bm \rho}-{\bm \Phi}_{\bf q {\bm \rho}}^{\rm T}{\alpha}\left(\dot{\bm \Phi}_{\bf q}\dot{\bf q}+
\dot{\bm \Phi}_{t}+2 \xi \omega \dot{\bm \Phi}+\omega^2 {\bm \Phi}\right)- \nonumber \\
&&{\bm \Phi}_{\bf q}^{\rm T}{\alpha}\left(\left(\dot{\bm \Phi}_{\bf q}\dot{\bf q}\right)_{\bm \rho}+
\dot{\bm \Phi}_{t{\bm \rho}}+2 \xi \omega \dot{\bm \Phi}_{\bm \rho}+\omega^2 {\bm \Phi}_{\bm \rho}\right), \\
\label{eq_dbarMdq_qs}
&&\bar{\bf M}_{\bf q} \ddot{\bf q}=
{\bf M}_{\bf q}\ddot{\bf q}+
{\bm \Phi}_{\bf qq}^{\rm T} \left( {\alpha}{\bm \Phi}_{\bf q}\ddot{\bf q} \right)+
{\bm \Phi}_{\bf q}^{\rm T} {\alpha} \left( {\bm \Phi}_{\bf qq}\ddot{\bf q} \right), \\
\label{eq_dbarMdro_qs}
&&\bar{\bf M}_{\bm \rho} \ddot{\bf q} =
{\bf M}_{\bm \rho}\ddot{\bf q}+
{\bm \Phi}_{\bf q {\bm \rho}}^{\rm T} \left( {\alpha} {\bm \Phi}_{\bf q}\ddot{\bf q} \right) +
{\bm \Phi}_{\bf q}^{\rm T}{\alpha} \left( {\bm \Phi}_{\bf q {\bm \rho}}\ddot{\bf q} \right).
\end{eqnarray}

In Eqns. \eqref{eq:barK} and \eqref{eq:barC}, ${\bf K} = -{\bf Q}_{\bf q}$ and ${\bf C} =-{\bf Q}_{\dot{\bf q}}$ respectively. For Eqns. \eqref{eq_dbarMdq_qs} and \eqref{eq_dbarMdro_qs}, the following magnitudes are tensor-vector products that have to be calculated as explained in the nomenclature
\begin{eqnarray}
&&{{\bf M}}_{\bf q} {\ddot{\bf q}} \equiv {{\bf M}}_{\bf q} \otimes {\ddot{\bf q}}\,, \\
&&{{\bf M}}_{\bm \rho} {\ddot{\bf q}} \equiv {{\bf M}}_{\bm \rho} \otimes {\ddot{\bf q}} \,, \\
&&{\bm \Phi}_{\bf qq}^{\rm T} \left( {\alpha}{\bm \Phi}_{\bf q}\ddot{\bf q} \right) \equiv {\bm \Phi}_{\bf qq}^{\rm T} \otimes \left( {\alpha}{\bm \Phi}_{\bf q}\ddot{\bf q} \right)\,, \\
&&{\bm \Phi}_{\bf q {\bm \rho}}^{\rm T} \left( {\alpha} {\bm \Phi}_{\bf q}\ddot{\bf q} \right) \equiv
{\bm \Phi}_{\bf q {\bm \rho}}^{\rm T} \otimes \left( {\alpha} {\bm \Phi}_{\bf q}\ddot{\bf q} \right)\,, \\
&&{\bm \Phi}_{\bf q}^{\rm T} {\alpha} \left( {\bm \Phi}_{\bf qq}\ddot{\bf q} \right) \equiv
{\bm \Phi}_{\bf q}^{\rm T} {\alpha} \left( {\bm \Phi}_{\bf qq} \otimes \ddot{\bf q} \right)\,, \\
&&{\bm \Phi}_{\bf q}^{\rm T}{\alpha} \left( {\bm \Phi}_{\bf q {\bm \rho}}\ddot{\bf q} \right) \equiv
{\bm \Phi}_{\bf q}^{\rm T}{\alpha} \left( {\bm \Phi}_{\bf q {\bm \rho}} \otimes \ddot{\bf q} \right)\,.
\end{eqnarray}

To obtain expression \eqref{eq:barK}, the kinematic relation \eqref{eq:dotPhi} was employed, and for expression \eqref{eq:barC} the relations $\left(\dot{\bm \Phi}_{\bf q}\right)_{\dot{\bf q}}={\bm \Phi}_{\bf qq}$, $\left(\dot{\bm \Phi}_{t}\right)_{\dot{\bf q}}={\bm \Phi}_{t {\bf q}}$, were used. The last two relations can be checked by considering the following differentials
\begin{eqnarray}
&&\delta{\bm \Phi}_{\bf q}={\bm \Phi}_{\bf qq} \delta{\bf q} \Rightarrow
\frac{d}{dt}\delta{\bm \Phi}_{\bf q}=
\dot{\bm \Phi}_{\bf qq} \delta{\bf q}+
{\bm \Phi}_{\bf qq} \delta\dot{\bf q}= \nonumber \\
&&\delta\dot{\bm \Phi}_{\bf q}=
\dot{\bm \Phi}_{\bf qq} \delta{\bf q}+
\dot{\bm \Phi}_{\bf q \dot{q}} \delta\dot{\bf q} \Rightarrow
\dot{\bm \Phi}_{\bf q \dot{q}}={\bm \Phi}_{\bf qq}\,, \\
&&\delta{\bm \Phi}_{t}={\bm \Phi}_{t{\bf q}} \delta{\bf q}
\Rightarrow \frac{d}{dt}\delta{\bm \Phi}_{t}=
\dot{\bm \Phi}_{t{\bf q}} \delta{\bf q}+
{\bm \Phi}_{t{\bf q}} \delta\dot{\bf q}= \nonumber \\
&&\delta\dot{\bm \Phi}_{t}=
\dot{\bm \Phi}_{t{\bf q}} \delta{\bf q}+
\dot{\bm \Phi}_{t\dot{\bf q}} \delta\dot{\bf q} \Rightarrow
\dot{\bm \Phi}_{t\dot{\bf q}}={\bm \Phi}_{t{\bf q}}\,.
\end{eqnarray}

\section{Sensitivity analysis of five-bar mechanism}
\begin{figure} [ht]
	\includegraphics[width=90mm]{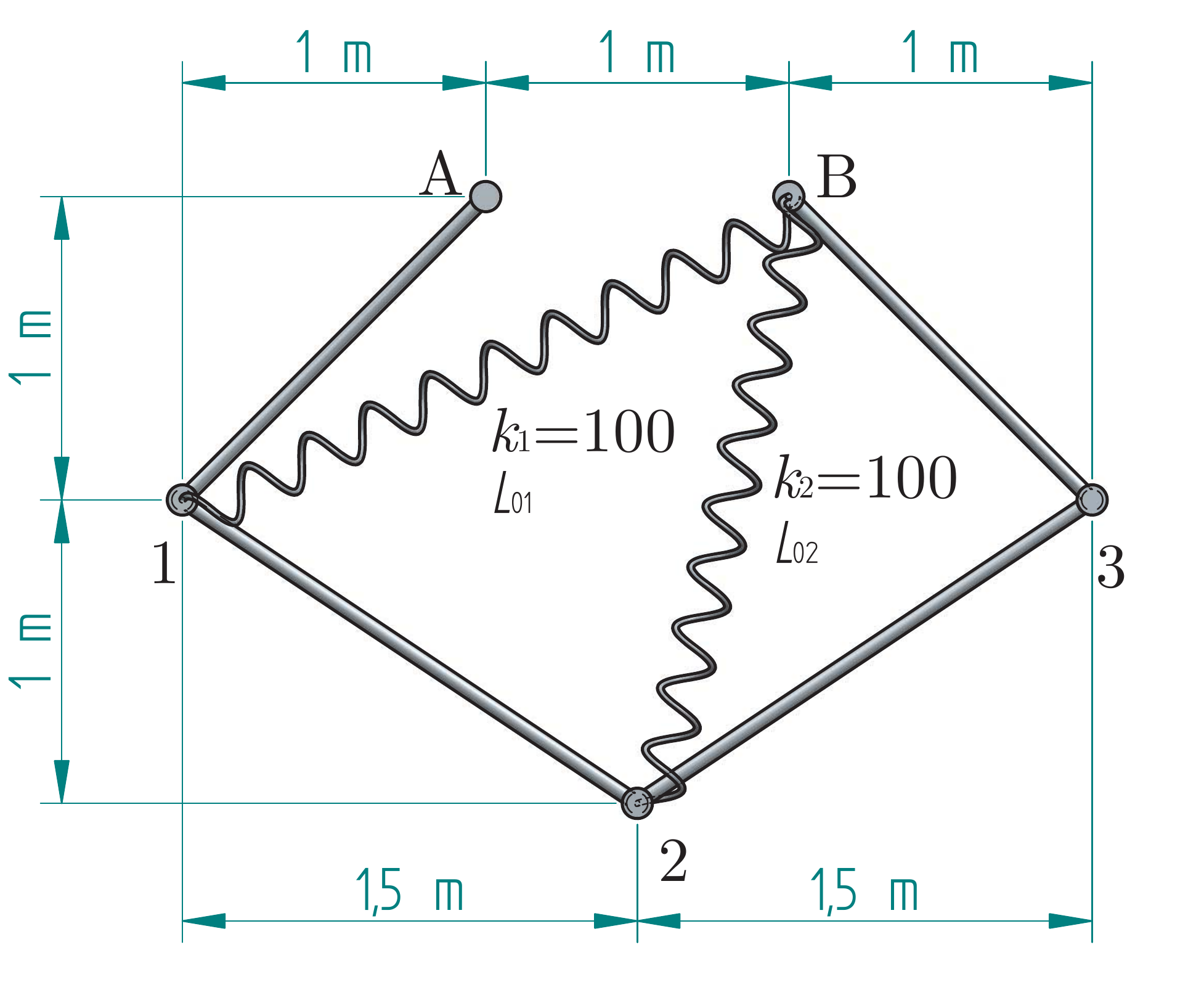}
	\caption{Five-bar mechanism}
	\label{Fig_fivebar}
\end{figure}

The mechanism chosen as a case study on which to test the sensitivity approach developed is the five-bar mechanism with two degrees of freedom shown in Fig.~\ref{Fig_fivebar}. The five bars are constrained by five revolute joints located at points A, 1, 2, 3, and B. The masses of the bars are $m_1=1 \; kg$, $m_2=1.5 \; kg$, $m_3=1.5 \; kg$, $m_4=1 \; kg$, and the polar moments of inertia are calculated under the assumption of a uniform distribution of mass. The mechanism is subjected to the action of gravity and of two elastic forces coming from the springs. The stiffness coefficients of the springs are $k_1=k_2=100 \; N/m$ and their natural lengths are initially chosen $L_{01}=\sqrt{2^2+1^2} \; m$ and $L_{02}=\sqrt{2^2+0.5^2} \; m$, coincident with the initial configuration shown in Fig.~\ref{Fig_fivebar}.

The following objective function is proposed to obtain its sensitivity with respect to the parameters ${\bm \rho}= \left[ \begin{array}{c c} L_{01} & L_{02} \end{array} \right]$.
\begin{equation}
\label{eq_cost_fivebar.old}
{\psi} = \int_{t_0}^{t_F} \left({\bf r}_2-{\bf r}_{20}\right)^{\rm T}
\left({\bf r}_2-{\bf r}_{20}\right) {\rm dt} \, ,
\end{equation}
where ${\bf r}_2$ is the global position of the point 2 and ${\bf r}_{20}$ is the initial position of the same point.

In order to validate the sensitivity approach derived in this paper, the sensitivities obtained for the five-bar mechanism are calculated using different approaches and compared.
The sensitivities, ${\nabla}_{\bm \rho}{\psi}$, were obtained by the following approaches:
\begin{enumerate}
\item Direct sensitivity approach using the index-3 DAEs formulation
\item Direct sensitivity approach using the index-1 DAEs formulation
\item Direct sensitivity approach using the penalty formulation
\item Direct sensitivity approach using Maggi's formulation
\item Adjoint sensitivity approach using the index-3 DAEs formulation
\item Adjoint sensitivity approach using the index-1 DAEs formulation
\item Adjoint sensitivity approach using the penalty formulation
\item Adjoint sensitivity approach using Maggi's formulation
\item Finite difference method with perturbation $\delta=10^{-7} m$.
\item Finite difference method with perturbation $\delta=10^{-4} m$.
\end{enumerate}

The results for the sensitivities with the mentioned approaches are presented in Table~\ref{table_results1}.

\begin{table}[ht]
\caption{Results for the five-bar mechanism.}
\begin{center}
\label{table_results1}
\begin{tabular}{l l l l}
& & &\\ 
\hline
Approach & Parameters & ${{\rm d}\psi}/{{\rm d}{L}_{01}}$ & ${{\rm d}\psi}/{{\rm d}{L}_{02}}$ \\
\hline
\hline
1: Direct index-3 & $h=10^{-2} s$ & -4.2381 & 3.2170 \\
2: Direct index-1 & $h=10^{-2} s$ & -4.2383 & 3.2169 \\
3: Direct penalty & $h=10^{-2} s$ & -4.2305 & 3.2154 \\
4: Direct Maggi's & $h=10^{-2} s$ & -4.2300 & 3.2112 \\
5: Adjoint index-3 & $h=10^{-2} s$ & -4.2287 & 3.2090 \\
6: Adjoint index-1 & $h=10^{-2} s$ & -4.2294 & 3.2094 \\
7: \textcolor{green}{Adjoint penalty} & $h=10^{-2} s$ & -4.2293 & 3.2137 \\
8: Ajoint Maggi's & $h=10^{-2} s$ & -4.2294 &  3.2093 \\
9: Num. diff. & $\delta=10^{-7} m$ & -9.7390 & -4.0344 \\
10: Num. diff. & $\delta=10^{-4} m$ & -4.2194 & 3.2055 \\
\hline
\hline
\end{tabular}
\end{center}
\end{table}

As it can be seen in Table~\ref{table_results1}, all the approaches employed yield similar results, as expected, except the finite difference method with perturbations $\delta=10^{-7} m$, thus perfectly validating the adjoint sensitivity approach using penalty formulation developed in this study. It is important to remark that, when the perturbation is too small, the sensitivities generated by finite difference method suffer from low accuracy due to round-off error generated by the computer. On the other hand, if the perturbation is too big, the finite difference method is not accurate anymore. Thus, under such circumstances, it is concluded that the finite difference method is not reliable. 

\section{Sensitivity analysis and optimization of the dynamic response of a full vehicle}
In addition to the five-bar mechanism, the approach derived in this paper has been used to perform the sensitivity analysis and the optimization of vehicle ride response for the Iltis vehicle shown in Fig. .~\ref{fig:Iltis}. The Iltis vehicle was proposed as a benchmark problem by the European automobile industry to check multibody dynamic codes. The vehicle model is extensively described in \cite{Iltis1990}, therefore only a summary of the most important parameters of the model is given here. Because the tire model employed in this study is different than the one described in the reference, it will be fully described here.

A speed bump test has been implemented. The objective function for this scenario is the fourth power of the chassis' \textit{CG} vertical acceleration. It is minimized while the vehicle goes straight over a road with a small step located at some distance ahead from the initial point. 

The sensitivities obtained for the vehicle model are not presented separately, but they are applied to perform the design optimization using by L-BFGS-B. The results for the optimization are presented.

\subsection{Iltis vehicle model}
\subsubsection{Vehicle topology}
\begin{figure}[h]
\centering
\includegraphics[width=90mm]{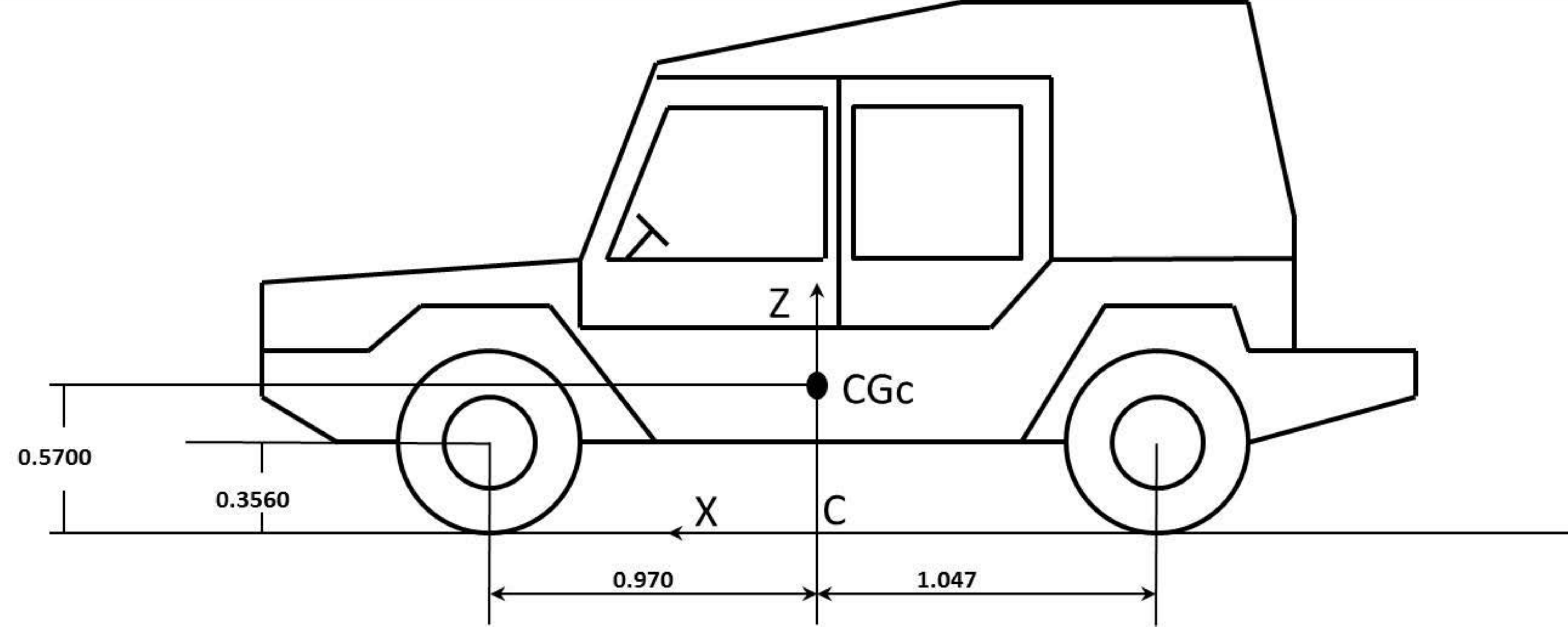}
\caption{The Bombardier Iltis vehicle}
\label{fig:Iltis}
\end{figure}
\begin{figure}[h]
\includegraphics[width=90mm]{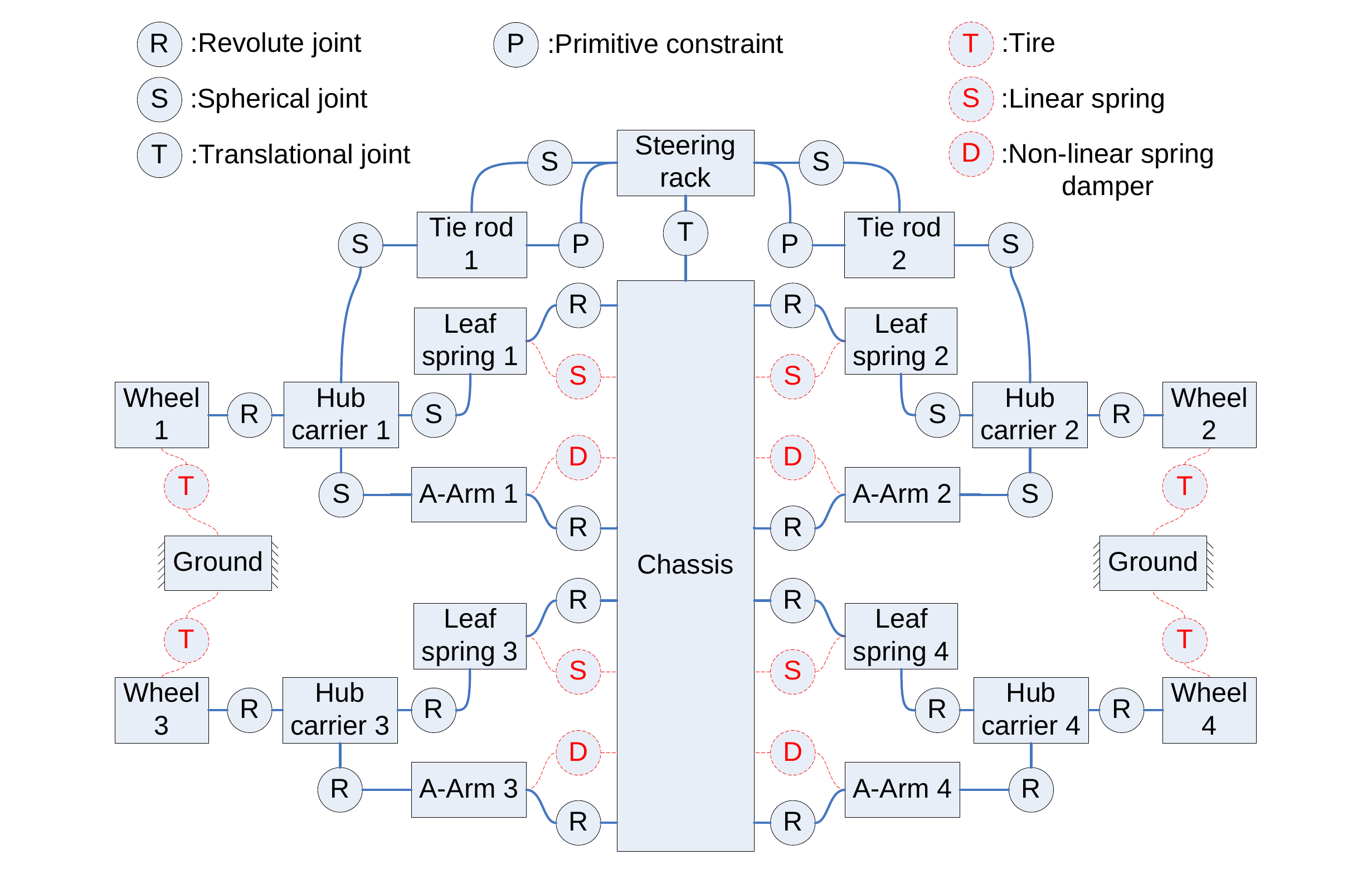}
\caption{Topology of the multibody vehicle model}
\label{fig:Iltis_diagram}
\end{figure}


The vehicle is represented in Fig.~\ref{fig:Iltis} and a topology diagram of the model is given in Fig.~\ref{fig:Iltis_diagram}, showing that the model is composed of 20 bodies: the chassis, 4 bodies per suspension, 1 tie rod per each one of the front suspensions, and the steering rod. The bodies of the model are joined by 25 kinematic joints plus 3 extra primitive constraints: 16 revolute joints, 8 spherical joints, 1 translational joint, 2 constraints to avoid the rotation of the tie rods, and a rheonomic constraint to control the steering rod. 
	
The total number of coordinates is 140 and the total number of constraints is 132 (6 of them redundant) giving a total count of 14 \textit{DOF}: 6 \textit{DOF} for the chassis, 4 \textit{DOF} for the suspensions and 4 \textit{DOF} for the wheels rotation. The steering is controlled by means of the mentioned rheonomic constraint and therefore it is not a true \textit{DOF} since it is kinematically determined.

Masses and moments of inertia are given in Table~\ref{table:mass_inertia}. As indicated in \cite{Iltis1990}, the masses of bodies not included in the table are neglected, and all the moments of inertia are principal, therefore they are given in their centroidal reference frames and all products of inertia are considered to be zero. Centers of mass locations are given in Table~\ref{table:COM}, expressed in the reference frame \textit{C}, indicated in Fig.~\ref{fig:Iltis}.

\begin{figure}[h]
\includegraphics[width=55mm]{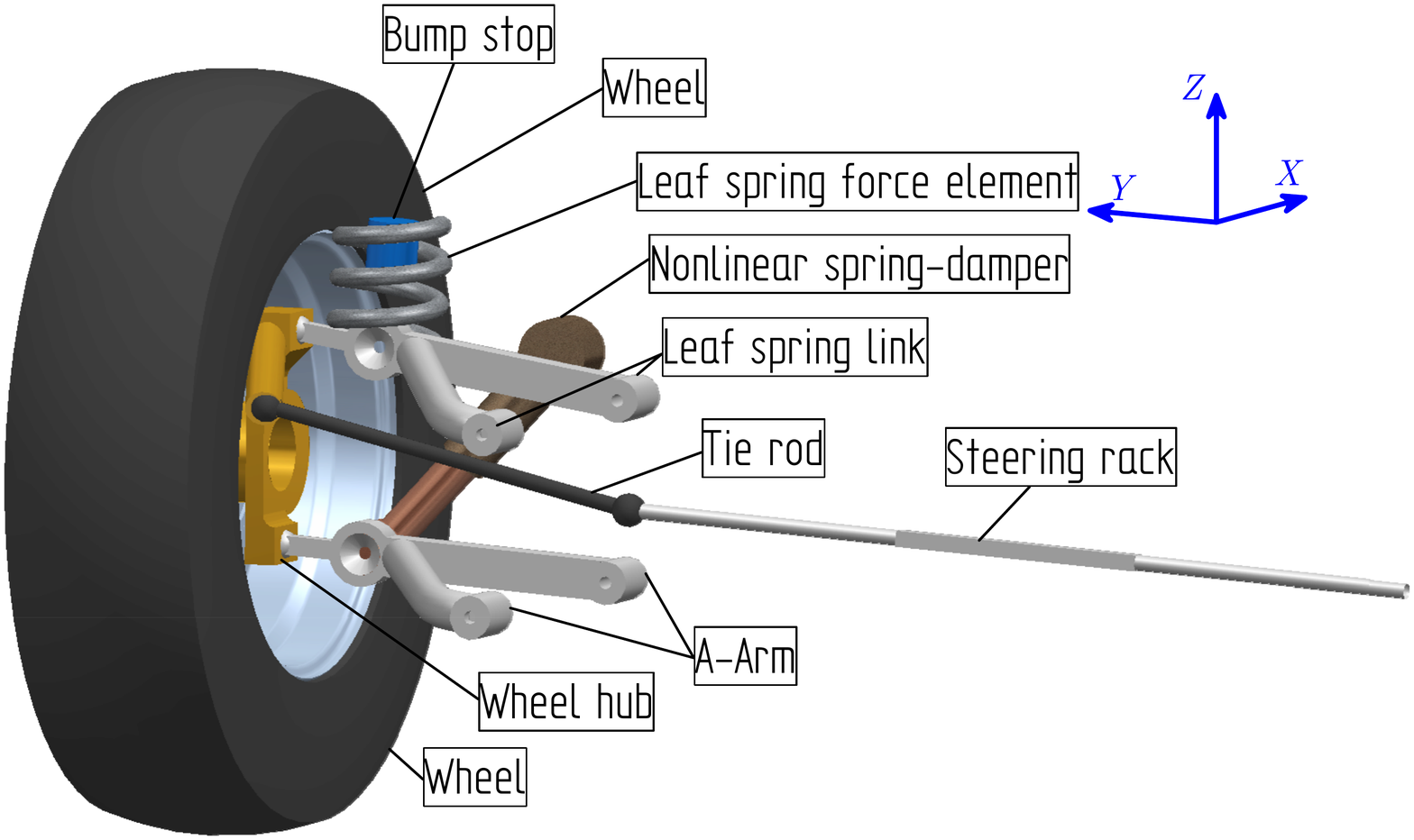}
\caption{Left front suspension system}
\label{fig:Iltis_frontsuspension}
\end{figure}

The geometry of the left front suspensions is shown in Fig.~\ref{fig:Iltis_frontsuspension}. The rear suspensions have a similar topology, but without the steering system. Note that the leaf spring is modeled as a link and a linear spring.
\\
\begin{table}[H]
\caption{Mass and principal moments of inertia}
\begin{center}
\label{table:mass_inertia}
\begin{tabular}{l l l l l}
\hline
\multirow{2}{*}{Body} &Mass&$I_{xx}$&$I_{yy}$&$I_{zz}$\\&$\ [\mathrm{kg}]$&$[\mathrm{kg\ m^2}]$&$[\mathrm{kg\ m^2}]$&$[\mathrm{kg\ m^2}]$ \\
\hline
\multirow{1}{0.9in}{Chassis} & \multirow{1}{0.4in}{1260} & \multirow{1}{0.4in}{130} & \multirow{1}{0.4in}{1620} & \multirow{1}{0.4in}{1670} \\
\hline
\multirow{2}{0.9in}{Wheel/hub/brake assembly} & \multirow{2}{0.4in}{57.35} & \multirow{2}{0.4in}{1.2402} & \multirow{2}{0.4in}{1.908} & \multirow{2}{0.4in}{1.2402}\\\\
\hline
\multirow{1}{0.9in}{A-arm} & \multirow{1}{0.4in}{6.0} & \multirow{1}{0.4in}{0.052099} & \multirow{1}{0.4in}{0.023235} & \multirow{1}{0.4in}{0.068864} \\
\hline
\end{tabular}
\end{center}
\end{table}
\begin{table}[H]
\caption{Positions of centers of mass (origin C. Fig.~\ref{fig:Iltis})}
\label{table:COM}
\begin{tabular}{l l l l}
\hline
\multirow{2}{*}{Body} & \multicolumn{3}{c}{Center of mass coordinates[m]}\\
\cline{2-4}
& $x$ & $y$ & $z$ \\
\hline
chassis & 0 & 0 & 0.57 \\
\hline
\multirow{2}{1.4in}{right front wheel with hub and brake assembly} & \multirow{2}{*}{0.97} & \multirow{2}{*}{-0.615} & \multirow{2}{*}{0.356} \\\\
\hline
\multirow{2}{1.4in}{left rear wheel with hub and brake assembly} & \multirow{2}{*}{-1.047} & \multirow{2}{*}{0.615} & \multirow{2}{*}{0.356} \\\\
\hline
\multirow{1}{*}{right front A-arm} & \multirow{1}{*}{0.97} & \multirow{1}{*}{-0.4155} & \multirow{1}{*}{0.2655} \\
\hline
\multirow{1}{*}{left rear A-arm} & \multirow{1}{*}{-1.047} & \multirow{1}{*}{0.4155} & \multirow{1}{*}{0.2655} \\
\hline
\end{tabular}
\end{table}

\subsubsection{The suspension model}

The key point positions for the left front suspension are given in Table~\ref{table:suspension_points}. The corresponding points for the left rear suspensions can be easily obtained, since all the suspensions are identical, except for the fact that the tie rods are not present in the rear, since there is no steering in the back.

\begin{table}[H]
\caption{Positions of joints (left front suspension, origin C. Fig.~\ref{fig:Iltis})}
\begin{center}
\label{table:suspension_points}
\begin{tabular}{p{1.75in} l l l}
\hline
Point description & x [m] & y [m] & z [m] \\
\hline
wheel center & 0.97 & 0.615 & 0.356 \\
A-arm to hub carrier & 0.97 & 0.572 & 0.229 \\
A-arm to chassis & 0.97 & 0.259 & 0.302 \\
leaf spring to hub carrier & 0.97 & 0.488 & 0.531 \\
leaf spring to chassis & 0.97 & 0.1585 & 0.600 \\
damper to A-arm & 1.045 & 0.500 & 0.241 \\
damper to chassis & 1.045 & 0.297 & 0.632 \\
tie rod to hub carrier & 0.83 & 0.448 & 0.531 \\
tie rod to chassis & 0.83 & 0.07 & 0.600 \\
steering rack to chassis & 0.83 & 0.00 & 0.600 \\
\hline
\end{tabular}
\end{center}
\end{table}

Each one of the four suspensions has three force elements: a linear leaf-spring that represents the stiffness of the leaf spring, a bump stop, and a non-linear spring-damper element. The suspension forces in the nominal configuration are given in Table~\ref{table_fsuspension}.
\begin{table}[H]
\caption{Suspension forces in the nominal configuration}
\begin{center}
\label{table_fsuspension}
\begin{tabular}{l l}
\hline
Leaf spring force & 2728.9 N \\
\hline
Non-linear Spring-Damper force & 128.0 N \\
\hline
Bump stop force & 0.0 N \\
\hline
\end{tabular}
\end{center}
\end{table}

The force of the leaf spring can be represented by the following equation,
\begin{equation}
\label{eq:leafspring}
F_L= -k_L \left(L-\left(1+2728.9/35906\ N/m\right)\right).
\end{equation}
where $L$ is distance between the spring extreme points, the stiffness is originally $k_L=35906\ N/m$, and in the nominal (initial) configuration $L=1\ m$ and the leaf spring force is equal to $F_L=2728.9\ N$.

The force of the bump stop is given by 

\begin{eqnarray}
\label{eq:bumpstop}
F_B =& -10^7 \left(s-0.93\right) \, &; \ s < 0.93\ m, \\
F_B =& 0 \, &; \ s \geq 0.93\  m.
\end{eqnarray}

The elastic and the damping forces of the nonlinear spring and damper system are given by the following expressions
\begin{eqnarray}
\label{suspe}
&&F_s=-4.0092\cdot10^6+k_S\cdot10^7 s \nonumber \\
&&-6.7061\cdot10^7 s^2+5.2796\cdot10^7 s^3, \\
&&F_d=c_S\cdot v+33955.72 v^2-59832.25 v^3 \nonumber \\
&&-395651.0 \, v^4; \ -0.2<v<0.21\ m/s, \\
&&F_d=-416.4200+1844.3v ; \ v<-0.2\ m/s, \\
&&F_d=1919.1638+1634.727v; \ v >0.21\ m/s. 
\end{eqnarray}
where $s$ is the distance between the extreme points of the nonlinear spring-damper, $v$ is the derivative of $s$, $c_S$ is the dominant damping coefficient, $k_S$ is the dominant stiffness coefficient, and originally $c_S=9945.627\ N\ s/m$, $k_S=2.8397\cdot10^7 N/m$.

\subsubsection{The tire model}
The tire model consists of normal, longitudinal, and lateral components. The normal tire model component is a linear spring-damper element, and the longitudinal and lateral models are linearized tire models with saturation. The normal tire model is 

\begin{equation}
{\bf F}_n = -k_n \left(r-R\right) {\bf n}; \ r<R.
\end{equation}
where $r$ is the distance from the center of the wheel to the ground, $R$ is the tire radius, and $\bf n$ is the normal vector to the ground in the center of the contact region. The normal tire forces in the nominal configuration are given in Table~\ref{table_ftire}.
\begin{table}[H]
\caption{Tire forces in the nominal configuration}
\begin{center}
\label{table_ftire}
\begin{tabular}{l l}
\hline
front tyre load & 3829.6 N \\
\hline
rear tyre load & 3593.6 N \\
\hline
\end{tabular}
\end{center}
\end{table}

The longitudinal and the lateral models implemented in this work are described in \cite{Luque2004}.
\begin{eqnarray}
\label{eq:Ftire}
&&{\bf F}_{t}=F_{x}{\bf b} +F_{y}\left({\bf n}\times{\bf b}\right), \\
\label{eq:Fxtire}
&&F_{x}= \left\{\begin{array}{cl}
\displaystyle\frac{{\mu}_x \left|{\bf F}_{rad}\right|}{\kappa_c} \kappa ; & \ \kappa \leq \kappa_c,
\\ {\mu}_x \left|{\bf F}_{rad}\right| ; & \ \kappa > \kappa_c, \end{array}\right.\\
\label{eq:Fytire}
&&F_{y}= \left\{\begin{array}{cl}
\displaystyle\frac{{\mu}_y \left|{\bf F}_{rad}\right|}{\alpha_c} \alpha ; & \ \alpha \leq \alpha_c,
\\ {\mu}_y \left|{\bf F}_{rad}\right| ; & \ \alpha > \alpha_c. \end{array}\right.
\end{eqnarray}
where ${\bf u}$ is the unit vector coincident with the wheel rotation axis, ${\bf b}=\left( {\bf u}\times{\bf n}\right)/{\left|{\bf u}\times{\bf n}\right|}$ is the longitudinal vector, $\kappa$ is the longitudinal slip, $\alpha$ is the slip angle, and $\kappa_c$, $\alpha_c$ are the critical slip factors for the longitudinal and lateral models, which are parameters of the tire model.

The longitudinal slip and the slip angle can be defined according to the following expressions, respectively
\begin{eqnarray}
&&\kappa = \displaystyle\frac{-{\bf b}^T{\bf v}_{slip}}{{\bf b}^T {\bf v}_c}= \displaystyle\frac{-{\bf b}^T({\bf v}_c-{\bf v}_{r})}{{\bf b}^T {\bf v}_c}= \nonumber \\
&&\displaystyle\frac{-{\bf b}^T\left({\bf v}_c-{\bf \omega}\times r{\bf n}\right)}{{\bf b}^T {\bf v}_c}, \\ 
&&\alpha= -\arcsin\left({\bf n}^T \left({\bf b} \times 
\displaystyle\frac{{\bf v}_{c}-\left({\bf n}^{\rm T}{\bf v}_{c}\right){\bf n}}
{\left|{\bf v}_{c}-\left({\bf n}^{\rm T}{\bf v}_{c}\right){\bf n}\right|}\right)\right).
\end{eqnarray}
where ${\bf v}_{c}$ is the velocity of the center of the wheel, $\bf \omega$ is the angular velocity of the wheel, and $r$ the effective radius defined before.

The saturation ellipse between the longitudinal and the lateral forces if given by the following expression
\begin{equation}
\label{eq:Tire_ellipse}
\left(\frac{F^{sat}_{x}}{\mu_x}\right)^2 +\left(\frac{F^{sat}_{y}}{\mu_y}\right)^2 \leq \left|{\bf F}_{rad}\right|^2.
\end{equation}
where $\mu_x$ and $\mu_y$ stand for the longitudinal and lateral friction coefficients and are parameters of the tire model.

If the components evaluated from Eqn.~\eqref{eq:Fxtire} and Eqn.~\eqref{eq:Fytire} are not inside the ellipse Eqn.~\eqref{eq:Tire_ellipse}, the saturation of the forces take place and the previously calculated forces Eqn.~\eqref{eq:Fxtire} and Eqn.~\eqref{eq:Fytire} don't hold. In this case they have to be replaced by the following
\begin{gather}
F^{sat}_{x}= \displaystyle\frac{\left|{\bf F}_{n}\right|}{\sqrt{\displaystyle\left(\frac{F_{x}}{\mu_x}\right)^2 +\left(\frac{F_{y}}{\mu_y}\right)^2}} F_{x} = \displaystyle\frac{\left|{\bf F}_{n}\right|}{f_{roz}} F_{x}, \\
F^{sat}_{y}= \displaystyle\frac{\left|{\bf F}_{n}\right|}{\sqrt{\displaystyle\left(\frac{F_{x}}{\mu_x}\right)^2 +\left(\frac{F_{y}}{\mu_y}\right)^2}} F_{y} = \displaystyle\frac{\left|{\bf F}_{n}\right|}{f_{roz}} F_{y}.
\end{gather}

\subsection{Dynamic response optimization}
\begin{figure} [H]
	\begin{center}
	\includegraphics[width=85mm]{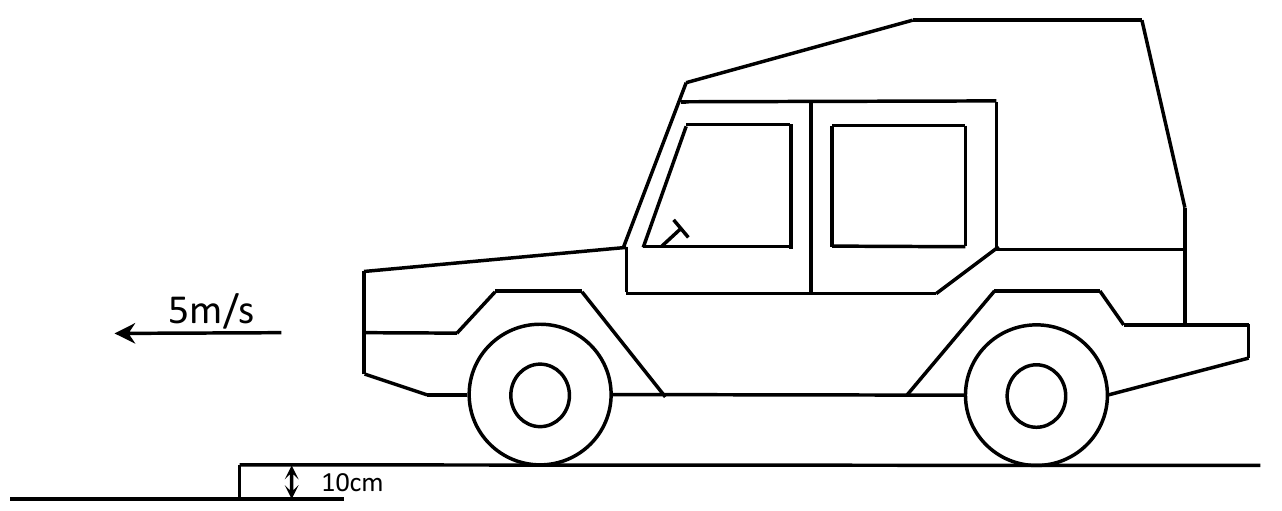}
	\end{center}
	\caption{The modified speed bumps test}
	\label{sbt}
\end{figure}
A speed bump test is implemented as a case study on which to perform design optimization. Typically, speed bumps are modeled as cylindrical shapes with the axis below the ground. However, the computation of the contacts between the bumps and the tires is complex. In order to simplify the speed bumps test, a modified speed bumps test is employed in this study which a step is used to replace those bumps as shown in Figure \ref{sbt}. At the beginning, the vehicle is released from equilibrium with an initial velocity of 5 m/s in the longitudinal direction. The steering is not actuated and the vehicle goes straight. At a distance of 6 m ahead from the initial position in the longitudinal direction, a step of 10 cm is placed. After 1 s the vehicle drops down the step and oscillates until the static equilibrium in the vertical direction is reached.

The objective function is the integral in time of the fourth power of the chassis \textit{CG} vertical acceleration
\begin{equation}
\psi =\int_{0}^{t}\; \ddot{z}_{\rm chassis}^4 \; dt\,.
\end{equation}

Six design parameters are chosen to perform design optimization. They are the stiffnesses of the rear and of the front   leaf springs $[k_{L1},k_{L2}]$ from \eqref{eq:leafspring}, the dominant damping coefficients of the rear and of the front suspension $[c_{S1},c_{S2}]$ from \eqref{suspe}, and the dominant stiffnesses of the rear and of the front suspensions $[k_{S1},k_{S2}]$ from \eqref{suspe}. The following constraints are imposed on the design parameters for the optimization problem:

\begin{eqnarray}
\label{ridecon}
&&0\le k_{L1},k_{L2},k_{S1},k_{S2} \le \infty \ N/m \\
&&0\le c_{S1},c_{S2} \le \infty \ N\ s/m 
\end{eqnarray}

\begin{figure}[H]
\includegraphics[width=80mm]{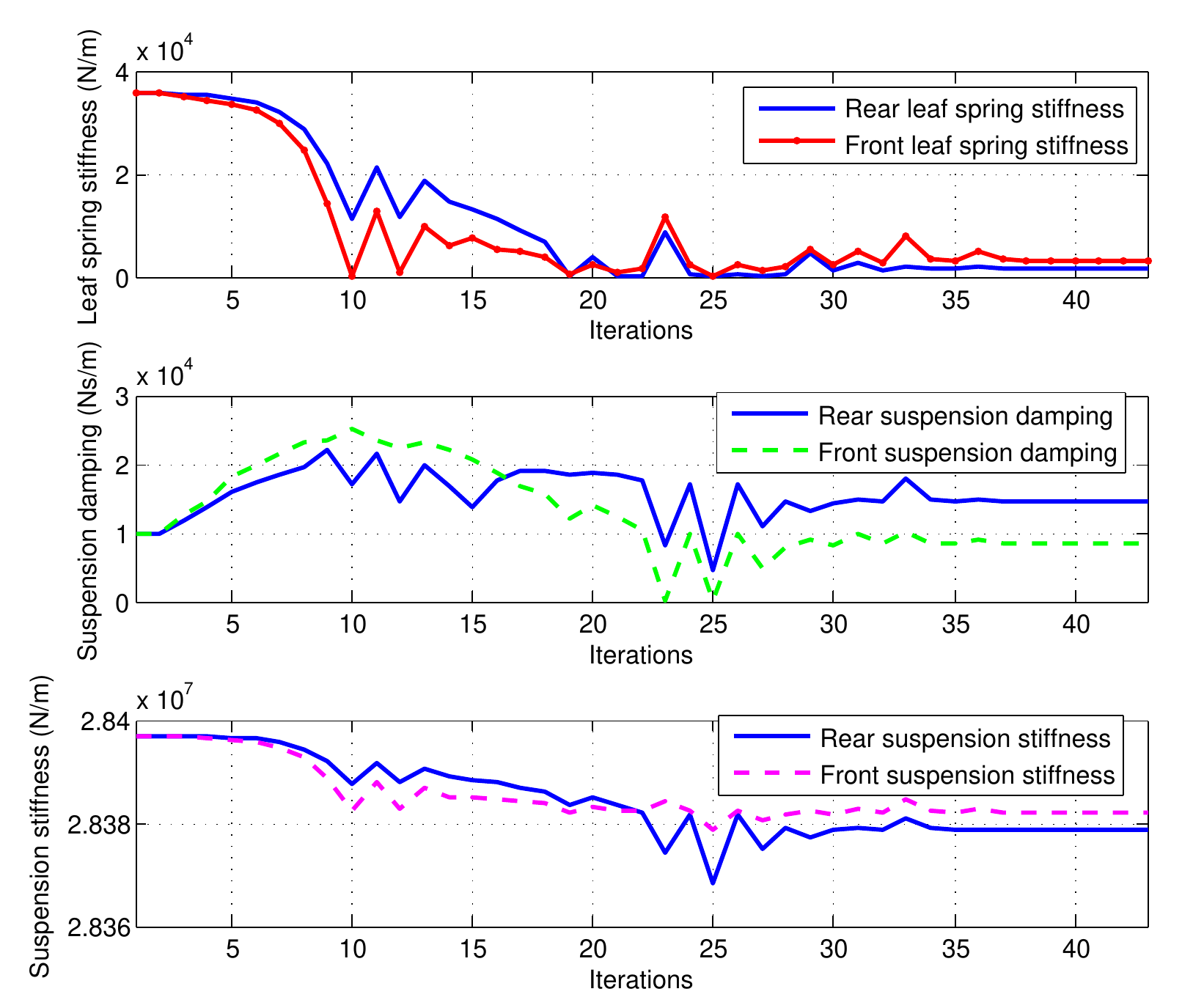}
\caption{The evolutions of the parameters of chassis vertical acceleration}
\label{fig:accel1}
\end{figure}

The evolutions of the parameters are given in Fig.~\ref{fig:accel1}, where it is shown that each parameter successfully converges after 40 iterations. The initial conditions of these parameters are the default values in \cite{Iltis1990}. The values of the non-optimized and of the optimized parameters and objective function are shown in Table \ref{table:op}.

\begin{figure}[H]
\includegraphics[width=80mm]{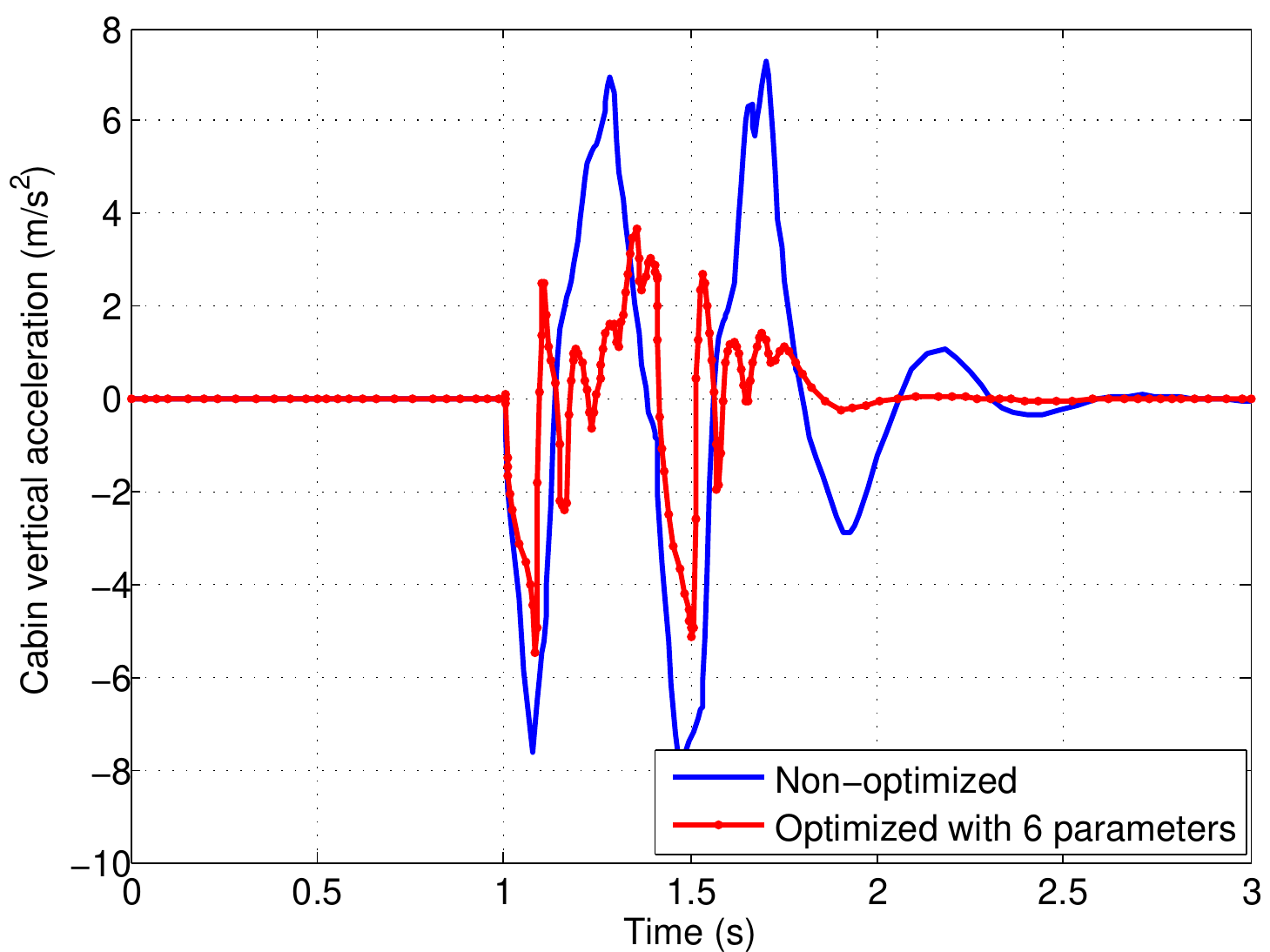}
\caption{Dynamic response of chassis vertical acceleration}
\label{fig:accel2}
\end{figure}

In Fig.~\ref{fig:accel2}, the dynamic responses of original and optimized systems are shown. It can be noted that the response is significantly improved. 

\begin{table}[H]
\caption{Optimized parameters and objective function}
\begin{center}
\label{table:op}
\begin{tabular}{l l l}
\hline
Optimizations & Non-optimized value & Optimized value\\
\hline
$k_{L1}$ [N/M] & 35906 & 1497\\
\hline
$k_{L2}$ [N/M] & 35906 & 3052\\
\hline
$c_{S1}$ [N s/M] & 9946 & 14531\\
\hline
$k_{L2}$ [N s/M] & 9946 & 8418\\
\hline
$k_{S1}$ [N/M] & 28397000 & 28378782\\
\hline
$k_{S2}$ [N/M] & 28397000 & 28382080\\
\hline
$\psi$ [$\rm N^4/s^7$] & 678 & 48\\
\hline
\end{tabular}
\end{center}
\end{table}

From the results above, it can be seen that the vehicle ride response is significantly improved by optimizing 6 parameters simultaneously; it is also shown that each parameter successfully converges after 40 iterations. From the gradient-based optimization point of view, the time of each iteration for optimization is highly related to the time needed to compute the sensitivities. If the sensitivity approach is computationally expensive, it takes a long time to finish each iteration, making gradient-based optimization very difficult. For the  optimization implemented in this section, it takes 406 seconds to finish the whole process in 43 iterations, which is very fast. Also, since 6 design parameters were chosen to perform the optimization, one can see that the adjoint sensitivity approach using the penalty formulation developed in this study is able to efficiently compute sensitivities for complex multibody systems with respect to multiple parameters.

\section{Conclusions}

This paper develops the theoretical adjoint sensitivity approach for multibody system dynamics based on penalty formulations, bringing new contributions to the state-of-the-art in analytical approaches for sensitivity analysis of such systems. A five-bar mechanism and a 14-\textit{DOF} vehicle model are implemented  as case studies to test and validate this sensitivity approach. For the five-bar mechanism, the sensitivity approach is validated by comparing the sensitivities generated using various sensitivity approaches. For the vehicle model, the optimization results presented clearly show that the vehicle ride response is significantly improved by optimizing 6 parameters; this demonstrates the capability of the new approach developed to perform sensitivity analysis for large and complex multibody systems with respect to multiple design parameters with accuracy and efficiency.

\begin{acknowledgment}
This work has been partially supported by NSF Award no. 1130667, by the Computational Science Laboratory, and by the Advanced Vehicle Dynamics Laboratory at Virginia Tech.
\end{acknowledgment}

%
% ...

\bibliographystyle{./asmems4}

\bibliography{./bibliography_JCND2013}

\begin{thebibliography}{10}

\bibitem{Haug1978}
Haug, E., and Arora, J., 1978.
\newblock ``Design sensitivity analysis of elastic mechanical systems''.
\newblock {\em Computer Methods in Applied Mechanics and Engineering, {\bf
  15}}, pp.~35--62.

\bibitem{Haug1984}
Haug, E.J., W.~R., and Mani, N., 1984.
\newblock ``Design sensitivity analysis of largescale constrained dynamic
  mechanical systems''.
\newblock {\em ASME Journal of Mechanisms, Transmissions, and Automation in
  Design, {\bf 106}}, pp.~156--162.

\bibitem{Bhatti1984}
Krishnaswami, P., and Bhatti, M., 1984.
\newblock ``A general approach for design sensitivity analysis of constrained
  dynamic systems''.
\newblock {\em ASME Journal of Mechanisms, Transmissions, and Automation in
  Design}, pp.~84--DET--132.

\bibitem{Haug1987}
Haug, E., 1987.
\newblock {\em Computer aided optimal design : structural and mechanical
  systems}.
\newblock No.~27 in NATO ASI series. Series F, Computer and systems sciences.
  Springer-Verlag, ch.~Design sensitivity analysis of dynamic systems.

\bibitem{chang1985}
Chang, C., and Nikravesh, P., 1985.
\newblock ``Optimal design of mechanical systems with constraint violation
  stabilization method''.
\newblock {\em Journal of Mechanisms, Transmissions and Automation in Design,
  {\bf 107}}(4), pp.~493--498.

\bibitem{Pagalday1997}
Pagalday, J., and Avello, A., {1997}.
\newblock ``{Optimization of multibody dynamics using object oriented
  programming and a mixed numerical-symbolic penalty formulation}''.
\newblock {\em Mechanism and Machine Theory, {\bf {32}}}({2}), {Feb},
  pp.~{161--174}.

\bibitem{haug1981design}
Haug, E., Wehage, R., and Barman, N., 1981.
\newblock ``Design sensitivity analysis of planar mechanism and machine
  dynamics''.
\newblock {\em Journal of Mechanical Design, {\bf 103}}(3), pp.~560--570.

\bibitem{Bestle1992}
Bestle, D., and Seybold, J., 1992.
\newblock ``Sensitivity analysis of constrained multibody systems''.
\newblock {\em Archive of Applied Mechanics, {\bf 62}}, pp.~181--190.

\bibitem{Bestle1992a}
Bestle, D., and Eberhard, P., 1992.
\newblock ``Analyzing and optimizing multibody systems''.
\newblock {\em Mechanics of Structures and Machines, {\bf 20}}(1), pp.~67--92.

\bibitem{Dias1997}
Dias, J., and Pereira, M., 1997.
\newblock ``Sensitivity analysis of rigid-flexible multibody systems''.
\newblock {\em Multibody System Dynamics, {\bf 1}}, pp.~303--322.

\bibitem{Feehery1997}
Feehery, W.~F., Tolsma, J.~E., and Barton, P.~I., 1997.
\newblock ``Efficient sensitivity analysis of large-scale
  differential-algebraic systems''.
\newblock {\em Applied Numerical Mathematics, {\bf 25}}(1), pp.~41 -- 54.

\bibitem{Anderson2002}
Anderson, K.~S., and Hsu, Y., 2002.
\newblock ``Analytical fully-recursive sensitivity analysis for multibody
  dynamic chain systems''.
\newblock {\em Multibody System Dynamics, {\bf 8}}, pp.~1--27.

\bibitem{Anderson2004}
Anderson, K., and Hsu, Y., 2004.
\newblock ``Order-(n+m) direct differentiation determination of design
  sensitivity for constrained multibody dynamic systems''.
\newblock {\em Structural and Multidisciplinary Optimization, {\bf 26}}(3-4),
  pp.~171--182.

\bibitem{Ding2007}
Ding, J.-Y., Pan, Z.-K., and Chen, L.-Q., 2007.
\newblock ``Second order adjoint sensitivity analysis of multibody systems
  described by differential-algebraic equations''.
\newblock {\em Multibody System Dynamics, {\bf 18}}, pp.~599--617.

\bibitem{Schaffer2006}
Schaffer, A., {2006}.
\newblock ``{Stabilized index-1 differential-algebraic formulations for
  sensitivity analysis of multi-body dynamics}''.
\newblock {\em Proceedings of the Institution of Mechanical Engineers Part K-
  Journal of Multi-Body Dynamics, {\bf {220}}}({3}), {SEP}, pp.~{141--156}.

\bibitem{Neto2009}
Neto, M.~A., Ambrosio, J. A.~C., and Leal, R.~P., 2009.
\newblock ``Sensitivity analysis of flexible multibody systems using composite
  materials components''.
\newblock {\em International Journal for Numerical Methods in Engineering, {\bf
  77}}(3), pp.~386--413.

\bibitem{Bhalerao2010}
Bhalerao, K., Poursina, M., and Anderson, K., 2010.
\newblock ``An efficient direct differentiation approach for sensitivity
  analysis of flexible multibody systems''.
\newblock {\em Multibody System Dynamics, {\bf 23}}, pp.~121--140.
\newblock 10.1007/s11044-009-9176-0.

\bibitem{Banerjee2013}
Banerjee, J.~M., and McPhee, J., 2013.
\newblock {\em Multibody Dynamics. Computational methods and applications.},
  Vol.~28 of {\em Computational Methods in Applied Sciences}.
\newblock Springer, ch.~Symbolic Sensitivity Analysis of Multibody Systems,
  pp.~123--146.

\bibitem{Brenan1989}
Brenan, K., Campbell, S., and Petzold, L., 1989.
\newblock {\em Numerical Solution of Initial-Value Problems in
  Differential-Algebraic Equations}.
\newblock North-Holland, New York.

\bibitem{Ascher1998}
Ascher, U., and Petzold, L., 1998.
\newblock {\em Computer methods for ordinary differential equations and
  differential-algebraic equations}.
\newblock Philadelphia Society for Industrial and Applied Mathematics.

\bibitem{dopicodirect}
Dopico, D., Zhu, Y., Sandu, A., and Sandu, C.
\newblock ``Direct and adjoint sensitivity analysis of {ODE} multibody
  formulations''.
\newblock {\em Journal of Computational and Nonlinear Dynamics}.

\bibitem{GarciadeJalon1994}
Garcia~de Jalon, J., and Bayo, E., 1994.
\newblock {\em Kinematic and dynamic simulation of multibody systems: The
  real-time challenge}.
\newblock Springer-Verlag, New York (USA).

\bibitem{jalon1994kinematic}
Jalon, J. G.~d., and Bayo, E., 1994.
\newblock ``Kinematic and dynamic simulation of multibody systems: the real
  time challenge''.

\bibitem{Bayo1988}
Bayo, E., Garc{\'\i}a~de Jalon, J., and Serna, M., 1988.
\newblock ``A modified {Lagrangian} formulation for the dynamic analysis of
  constrained mechanical systems''.
\newblock {\em Computer Methods in Applied Mechanics and Engineering, {\bf
  71}}(2), 11, pp.~183--195.

\bibitem{lbfgsb}
Zhu, C., Byrd, R.~H., Lu, P., and Nocedal, J., 1997.
\newblock ``Algorithm 778: {L-BFGS-B}: Fortran subroutines for large-scale
  bound-constrained optimization''.
\newblock {\em ACM Trans. Math. Softw., {\bf 23}}(4), Dec., pp.~550--560.

\bibitem{Cao2003}
Cao, Y., Li, S., Petzold, L., and Serban, R., 2003.
\newblock ``Adjoint sensitivity analysis for differential-algebraic equations:
  The adjoint {DAE} system and its numerical solution''.
\newblock {\em SIAM Journal on Scientific Computing, {\bf 24}}({3}),
  pp.~1076--1089.

\bibitem{Iltis1990}
Frik, S., Leister, G., and Schwartz, W., 1993.
\newblock ``Simulation of the {IAVSD} road vehicle benchmark bombardier iltis
  with {FASIM}, {MEDYNA}, {NEWEUL} and {SIMPACK}''.
\newblock {\em Vehicle System Dynamics, {\bf 22}}(sup1), pp.~215--253.

\bibitem{Luque2004}
Pablo Luque~Rodríguez, Daniel Álvarez~Mántaras, C.~V., 2004.
\newblock {\em Ingeniería del automóvil: sistemas y comportamiento
  dinámico}.
\newblock THOMSON.

\end{thebibliography}



\end{document}